\documentclass{aa}  

\usepackage{graphicx}
\usepackage{txfonts}
\begin{document}

   \title{Precursors of fatty alcohols in the ISM: Discovery of {\it n}-propanol}

\authorrunning{Jim\'enez-Serra et al.}
\titlerunning{Fatty alcohols in the ISM}

   \author{Izaskun Jim\'enez-Serra
          \inst{1},
          Lucas F. Rodr\'{\i}guez-Almeida\inst{1},
          Jes\'us Mart\'{\i}n-Pintado\inst{1}, V\'{\i}ctor M. Rivilla\inst{1,2}, Mattia Melosso\inst{3},  Shaoshan Zeng\inst{4}, Laura Colzi\inst{1,2},  Yoshiyuki Kawashima\inst{5}, Eizi Hirota\inst{6}, Cristina Puzzarini\inst{3}, Bel\'en Tercero\inst{7}, Pablo de Vicente\inst{7}, Fernando Rico-Villas\inst{1}, Miguel A. Requena-Torres\inst{8,9} \and
          Sergio Mart\'{\i}n\inst{10,11}
%
          }

   \institute{Centro de Astrobiolog\'{\i}a (CSIC/INTA), Ctra. de Torrej\'on a Ajalvir km 4, E-28806, Torrej\'on de Ardoz, Spain\\       \email{ijimenez@cab.inta-csic.es}
   \and
INAF-Osservatorio Astrofisico di Arcetri, Largo Enrico Fermi 5, 50125, Florence, Italy
\and
Dipartimento di Chimica "Giacomo Ciamician", Universit\'a di Bologna, via F. Selmi 2,40126, Bologna, Italy
\and
Star and Planet Formation Laboratory, Cluster for Pioneering Research, RIKEN,2-1 Hirosawa, Wako, Saitama, 351-0198, Japan
\and
Department of Applied Chemistry, Faculty of Engineering, Kanagawa Institute of Technology, Atsugi, Kanagawa 240-0292, Japan
\and
The Graduate University for Advanced Studies,
Hayama, Kanagawa 240-0193, Japan
\and
Observatorio de Yebes (IGN), Cerro de la Palera s/n, E-19141, Guadalajara, Spain
\and
University of Maryland, College Park, ND 20742-2421, USA 
\and
Department of Physics, Astronomy and Geosciences, Towson University, MD 21252, USA
\and
European Southern Observatory, Alonso de C\'ordova 3107, Vitacura 763 0355, Santiago, Chile
\and
Joint ALMA Observatory, Alonso de C\'ordova 3107, Vitacura 763 0355, Santiago, Chile
             }

   \date{Received September 15, 1996; accepted March 16, 1997}

 
  \abstract
   {Theories on the origins of life propose that early cell membranes were synthesized from amphiphilic molecules simpler than phospholipids such as fatty alcohols. The discovery in the interstellar medium (ISM) of ethanolamine, the simplest phospholipid head group, raises the question whether simple amphiphilic molecules are also synthesized in space.} 
   {We investigate whether precursors of fatty alcohols are present in the ISM.}
   {For this, we have carried out a spectral survey at 7, 3, 2 and 1$\,$mm toward the Giant Molecular Cloud G+0.693-0.027 located in the Galactic Center using the IRAM 30$\,$m and Yebes 40$\,$m telescopes.}
   {Here, we report the detection in the ISM of the primary alcohol {\it n}-propanol (in both conformers {\it Ga}-{\it n}-C$_3$H$_7$OH and {\it Aa}-{\it n}-C$_3$H$_7$OH), a precursor of fatty alcohols. The derived column densities of {\it n}-propanol are (5.5$\pm$0.4)$\times$10$^{13}$$\,$cm$^{-2}$ for the {\it Ga} conformer and (3.4$\pm$0.3)$\times$10$^{13}$$\,$cm$^{-2}$ for the {\it Aa} conformer, which imply molecular abundances of (4.1$\pm$0.3)$\times$10$^{-10}$ for {\it Ga}-{\it n}-C$_3$H$_7$OH and of (2.5$\pm$0.2)$\times$10$^{-10}$ for {\it Aa}-{\it n}-C$_3$H$_7$OH. We also searched for the {\it AGa} conformer of n-butanol [{\it AGa}-{\it n}-C$_4$H$_9$OH] without success yielding an upper limit to its abundance of $\leq$4.1$\times$10$^{-11}$. The inferred CH$_3$OH:C$_2$H$_5$OH:C$_3$H$_7$OH:C$_4$H$_9$OH abundance ratios go as 1:0.04:0.006:$\leq$0.0004 toward G+0.693-0.027, i.e. they decrease roughly by one order of magnitude for increasing complexity. We also report the detection of both {\it syn} and {\it anti} conformers of vinyl alcohol, with column densities of (1.11$\pm$0.08)$\times$10$^{14}$$\,$cm$^{-2}$ and (1.3$\pm$0.4)$\times$10$^{13}$$\,$cm$^{-2}$, and abundances of (8.2$\pm$0.6)$\times$10$^{-10}$ and (9.6$\pm$3.0)$\times$10$^{-11}$, respectively.}
   {The detection of {\it n}-propanol, together with the recent discovery of ethanolamine in the ISM, opens the possibility that precursors of lipids according to theories of the origin of life, could have been brought to Earth from outer space.}

\keywords{ISM: molecules -- ISM: clouds -- Galaxy: centre -- astrochemistry -- line: identification}

\maketitle
%

\section{Introduction}

Compartmentalization (i.e. the formation of vesicles or small membranes) is an essential step in the emergence of life. Theories of the origin of life such as prebiotic systems chemistry \citep[][]{ruiz-mirazo14}, propose that compartmentalization had to appear concurrently with the first metabolic and replicative systems forming proto-cells. It is believed that the composition of primordial membranes was much simpler than that of modern cells. Several families of amphiphilic molecules such as alkyl phosphates, alkyl sulfates, fatty acids, and fatty alcohols have been proposed as possible prebiotic lipids, i.e. as constituents of proto-membranes \citep[][]{ruiz-mirazo14}. Among them, fatty alcohols (also called primary alcohols) present a simple structure consisting of a straight-chain with the OH group linked to a primary carbon (i.e. forming a -CH$_2$OH group). 

Two possible scenarios have been proposed for the formation of prebiotic lipids: they could form {\it endogenously} under early Earth conditions via e.g. Fischer-Tropsch synthesis \citep[][]{nooner76,mccollom99,ruiz-mirazo14}; or they could be synthesised {\it exogenously} in interstellar space and incorporated to Solar-system materials such as interplanetary dust particles, comets and meteorites \citep[][]{bernstein99}. Although not all examined meteorites contain long-chain amphiphiles \citep{deamer94}, \citet{deamer85} showed that the amphiphilic compounds found in the Murchison meteorite can form vesicle-like structures under certain prebiotic conditions. Using the different contributions to organic matter delivery analysed by \citet[][see their Table$\,$2]{chyba92}, \citet{ruiz-mirazo14} estimated that a total amount of $\sim$10$^{16}$-10$^{18}$$\,$kg of extraterrestrial organic matter was delivered to Earth during the Late Heavy Bombardment period (mainly between 3.9 and 3.8 billions of years ago), which supports the idea that extraterrestrial material could have contributed to the emergence of prebiotic lipids in an early Earth\footnote{Note that the total amount of organic matter in the present biosphere is about 6$\times$10$^{14}$$\,$kg \citep[][]{bar-on18}.}.

The recent discovery of ethanolamine (NH$_2$CH$_2$CH$_2$OH), the simplest head group of phospholipids \citep[][]{rivilla21a}, in the interstellar  medium (ISM), has raised the interest for the search of precursors of prebiotic lipids in space, and in particular of other alcohols. For fatty alcohols, the only ones that have been observed in the ISM are methanol (CH$_3$OH) and ethanol (C$_2$H$_5$OH), which are practically ubiquitous in the ISM \citep[see e.g.][and references therein]{jorgensen20,martin21}. {\it n}-propanol (also named 1-propanol; molecular formula CH$_3$CH$_2$CH$_2$OH or C$_3$H$_7$OH) has been searched for in the ISM, but no detection has been reported to date  \citep[][]{tercero15,muller16,qasim19,manigand21}. However, primary alcohols, including propanol, have been detected in comet 67P/Churyumov-Gerasimenko \citep[][]{altwegg19}, which suggests that this molecule could be formed under interstellar conditions.  

In this work, we report the discovery of {\it n}-propanol toward the Giant Molecular Cloud (GMC) G+0.693-0.027 (hereafter G+0.693) located in the Galactic Center. We also report the detection of both {\it syn} and {\it anti} conformers of vinyl alcohol (CH$_2$CHOH) toward the same cloud. The {\it syn} conformer was tentatively detected toward SgrB2(N) by \citet{turner01}, but this detection was never confirmed in higher-sensitivity observations carried out with ALMA \citep[][]{martin-drumel19}. More recently, \citet{agundez21} have unambiguously identified  the {\it syn} conformer toward TMC-1. However, to our knowledge this is the first time that the {\it anti} conformer of this molecule is robustly detected in the ISM. 

G+0.693 is known to be one of the most chemically-rich interstellar sources in our Galaxy \citep[][]{requena-torres06, requena-torres08, zeng18}, despite the fact it does not show any sign of high-mass star-formation activity \citep[][]{zeng20}. The chemistry of this cloud is characterized by the presence of low-velocity shocks \citep[its emission shows broad linewidths of $\sim$20$\,$km$\,$s$^{-1}$ and high abundances of HNCO, a typical molecular tracer of low-velocity shocks;][]{martin08}, which sputter the molecular content of the icy mantles of dust grains into the gas phase \citep[see e.g.][]{jimenez-serra08}. The H$_2$ gas densities in G+0.693 are of the order of a few 10$^4$$\,$cm$^{-3}$ \citep[][]{zeng20} and the gas kinetic temperatures range from 70 to 150$\,$K, as measured using CH$_3$CN \citep[][]{zeng18}. Due to the low H$_2$ gas densities, the molecular line emission from high dipole moment molecules such as complex organics is sub-thermally excited, and their excitation temperature lies below 20$\,$K \citep[][]{requena-torres06,zeng18}. This represents an advantage for the search of new molecular species since the millimeter spectra observed toward this source present rather low levels of line blending and line confusion. As a result, G+0.693 has yielded several first detections of new molecular species in the ISM including {\it Z}-cyanomethanimine \citep[Z-HNCHCN;][]{rivilla19}, hydroxylamine \citep[NH$_2$OH;][]{rivilla20}, ethanolamine \citep[NH$_2$CH$_2$CH$_2$OH;][]{rivilla21a}, cyanomydil radical \citep[HNCN;][]{rivilla21b}, mono-thioformic acid \citep[HC(O)SH;][]{rodriguez-almeida21a}, ethyl isocyanate \citep[C$_2$H$_5$NCO;][]{rodriguez-almeida21b} and PO$^+$ \citep[][]{rivilla22}. Urea, an important prebiotic molecule \citep[][]{becker19,belloche19}, has also been detected toward this cloud \citep[][]{jimenez-serra20}.  

The paper is organised as follows. In Section$\,$2, we describe the observations, and in Section$\,$3 we present the detections of {\it n}-propanol and the {\it syn} and {\it anti} forms of vinyl alcohol. We also report the upper limits to the non-detections of isopropanol (i-C$_3$H$_7$OH), ethyl methyl ether (CH$_3$OCH$_2$CH$_3$), cyclopropanol (c-C$_3$H$_5$OH) and n-butanol (n-C$_4$H$_9$OH). In Section$\,$4, we discuss the chemistry of {\it n}-propanol and vinyl alcohol and put these detections in context with previous works and with theories for the origin of life. Finally, in Section$\,$5 we present our conclusions. 

\begin{table*}
\caption{Unblended or slightly blended transitions of {\it Ga}-{\it n}-C$_3$H$_7$OH and {\it Aa}-{\it n}-C$_3$H$_7$OH detected towards G+0.693.}
\scriptsize
\centering
\begin{tabular}{ccccccccc}
\hline\hline
Frequency & QNs
& E$_{\mathrm{u}}$ &$\log\mathrm{I(300K)}$ &rms& $\delta v$ & $\int T_A^* d\nu$&S/N\tablefootmark{a}&blending\\
(MHz)&(J$''_{\mathrm{K_a'',K_c''}}$- J$'_{\mathrm{K_a',K_c'}}$)& (K) &(nm$^{2}$$\,$MHz)&(mK)& ($\mathrm{km\,s^{-1}}$) & ($\mathrm{mK\,km\,s^{-1}}$) & & \\ \hline
\multicolumn{9}{c}{{\bf {\it Ga}-{\it n}-C$_3$H$_7$OH}} \\ \hline
40341.1539  & 3$_{1,2}$-2$_{0,2}$     & 3.3 & -6.4573 & 1.1 & 1.9 & 28.9(6.5)\tablefootmark{b} & 4.5 & blended with C$_2$H$_5$NCO \\ 
88449.6240   & 7$_{1,6}$-6$_{0,6}$     & 13.6 & -5.6188 & 1.1 & 4.2 & 47.5(9.9) & 4.8 & clean transition \\ 
94895.9659  & 5$_{3,2}$-4$_{2,2}$     & 11.0 & -5.5443 & 3.2 & 1.9 & 63.2(20.0) & 3.2 &  blended with HCCCC$^{13}$CN \\ 
97219.5786  & 11$_{0,11}$-10$_{1,10}$ & 28.8 & -5.0046 & 2.4 & 3.8 & 63.6(20.8) & 3.1 & clean transition \\ 
105032.7600\tablefootmark{*}   & 4$_{4,1}$-3$_{3,0}$     & 11.9 & -5.3260 & 2.2 & 1.7 & 335.6(13.0)\tablefootmark{c} & 25.7\tablefootmark{c} & clean transition\tablefootmark{c,d} \\ 
105032.8215\tablefootmark{*} & 4$_{4,0}$-3$_{3,0}$     & 11.9 & -5.3814 & 2.2 & 1.7 & -  & - & - \\
105034.6719\tablefootmark{*} & 4$_{4,1}$-3$_{3,1}$     & 11.9 & -5.3814 & 2.2 & 1.7 & - & - & - \\
105034.7334\tablefootmark{*} & 4$_{4,0}$-3$_{3,1}$     & 11.9  & -5.3259 & 2.2 & 1.7 & - & - & - \\
133685.6682\tablefootmark{*} & 5$_{5,0}$-4$_{4,1}$     & 18.3 & -5.0148 & 4.1 & 7.3 & 354.9(50.0)\tablefootmark{c} & 7.1\tablefootmark{c} & slightly blended with CH$_3$COCH$_3$\tablefootmark{c,d} \\ 
133685.6682\tablefootmark{*} & 5$_{5,1}$-4$_{4,0}$     & 18.3 & -5.0148 & 4.1 & 7.3 & - & - & - \\
133685.6682\tablefootmark{*} & 5$_{5,1}$-4$_{4,1}$     & 18.3 & -5.0718 & 4.1 & 7.3 & - & - & - \\
133685.6682\tablefootmark{*} & 5$_{5,0}$-4$_{4,0}$     & 18.3 & -5.0718 & 4.1 & 7.3 & - & - & - \\
143143.8136\tablefootmark{*} & 6$_{5,2}$-5$_{4,1}$     & 21.1 & -4.9557 & 1.5 & 1.7 & 320.2(8.5)\tablefootmark{c} & 37.7\tablefootmark{c} & clean transition\tablefootmark{c,d} \\
143143.8332\tablefootmark{*} & 6$_{5,1}$-5$_{4,1}$     & 21.1 & -5.0037 & 1.5 & 1.7 & - & - & - \\
143144.1096\tablefootmark{*} & 6$_{5,1}$-5$_{4,2}$     & 21.1 & -4.9557 & 1.5 & 1.7 & - & - & - \\
143144.1096\tablefootmark{*} & 6$_{5,2}$-5$_{4,2}$     & 21.1 & -5.0037 & 1.5 & 1.7 & -  & - & - \\
162029.4986\tablefootmark{*} & 8$_{5,4}$-7$_{4,3}$     & 27.9 & -4.8415 & 2.1 & 6.0 & 238.8(22.7)\tablefootmark{c} & 10.5\tablefootmark{c} & slightly blended with CH$_3$CHO\tablefootmark{c,d} \\
162030.2797\tablefootmark{*} & 8$_{5,3}$-7$_{4,3}$     & 27.9 & -4.8688  & 2.1 & 6.0 & - & - & - \\
162039.5868\tablefootmark{*} & 8$_{5,4}$-7$_{4,4}$     & 27.9 & -4.8687 & 2.1 & 6.0 & - & - & - \\
162040.3663\tablefootmark{*} & 8$_{5,3}$-7$_{4,4}$     & 27.9 & -4.8415 & 2.1 & 6.0 & - & - & - \\
162332.1928\tablefootmark{*} & 6$_{6,0}$-5$_{5,1}$     & 26.1 & -4.7688 & 4.2 & 6.0 & 307.0(46.0)\tablefootmark{c} & 6.7\tablefootmark{c} & slightly blended with HSC\tablefootmark{c} \\ 
162332.1928\tablefootmark{*} & 6$_{6,1}$-5$_{5,0}$     & 26.1 & -4.7688 & 4.2 & 6.0 & - & - & - \\
162332.1928\tablefootmark{*} & 6$_{6,1}$-5$_{5,1}$     & 26.1 & -4.8267 & 4.2 & 6.0 & - & - & - \\
162332.1928\tablefootmark{*} & 6$_{6,0}$-5$_{5,0}$     & 26.1 & -4.8267 & 4.2 & 6.0 & - & - & - \\ \hline
\multicolumn{9}{c}{{\bf {\it Aa}-{\it n}-C$_3$H$_7$OH}} \\ \hline
34110.7596  & 12$_{1,11}$-12$_{0,12}$ & 29.1 & -5.4467 & 0.9 & 2.2 & 33.6(6.0)\tablefootmark{b} & 5.6 & clean transition\\ 
44022.8732  & 3$_{1,3}$-2$_{0,2}$     & 3.2 & -5.8803 & 2.2 & 1.7 & 71.0(12.8) & 5.5\tablefootmark{e} & clean transition \\ 
73921.2560   & 9$_{2,8}$-9$_{1,9}$     & 20.2 & -5.1140  & 2.8 & 2.5 & 69.2(20.0) & 3.5 & blended with U species \\ 
77153.9362\tablefootmark{*}  & 8$_{1,8}$-7$_{0,7}$     & 13.6 & -5.0315 & 1.4 &  2.4 & 134.2(9.3) & 14.4 & slightly blended with n-C$_3$H$_7$CN\tablefootmark{d}\\ 83485.2943\tablefootmark{*}  & 9$_{1,9}$-8$_{0,8}$     & 16.7 & -4.9139 & 2.3 & 2.2 & 128.9(15.0) & 8.6 & clean transition\tablefootmark{d} \\
89742.8720\tablefootmark{*}   & 10$_{1,10}$-9$_{0,9}$   & 20.2 & -4.8060  & 1.6 & 2.0 & 118.6(10.3) & 11.5 & slightly blended with (CH$_2$OH)$_2$ and U-line\tablefootmark{d}\\ 
89850.2691\tablefootmark{*}  & 3$_{2,2}$-2$_{1,1}$     & 6.5 & -5.3442 & 1.3 & 2.0 & 98.0(8.3) & 11.8 & clean transition\tablefootmark{d} \\
90618.9845\tablefootmark{*}  & 3$_{2,1}$-2$_{1,2}$     & 6.5 & -5.3441 & 2.4 & 2.0 & 97.3(15.3) & 6.4 & slightly blended with HCC$^{13}$CN \\ 
150358.3345\tablefootmark{*} & 5$_{3,3}$-4$_{2,2}$     & 15.1 & -4.6853 & 2.9 & 1.6 & 144.2(16.6) & 8.7 & clean transition \\
150390.2490\tablefootmark{*}  & 5$_{3,2}$-4$_{2,3}$     & 15.1 & -4.6852 & 2.9 & 1.6 & 144.2(16.6) & 8.7 & clean transition \\ \hline
\end{tabular}
\tablefoot{
\tablefoottext{a}{The S/N ratios are calculated from the integrated intensity of the lines inferred from the LTE fits of MADCUBA-SLIM. Note that they are not calculated from the gaussian fits of individual lines.}
\tablefoottext{b}{The error in the integrated intensity is calculated as $\rm rms$$\times$$\sqrt{\delta \rm v \times \Delta \rm v}$, with $\delta \rm v$ the velocity resolution of the spectra and $\Delta \rm v$ the FWHM of the line emission of 20$\,$km$\,$s$^{-1}$.}
\tablefoottext{c}{The line corresponds to a blend of four individual {\it Ga}-{\it n}-C$_3$H$_7$OH transitions. The total integrated intensity and the S/N ratio are calculated from the sum of the areas of the individual transitions.}
\tablefoottext{d}{Transitions detected at the $\geq$3$\,$rms noise level in peak intensity.}
\tablefoottext{e}{This S/N ratio should be taken with caution since the LTE fit carried out by MADCUBA-SLIM overpredicts the integrated intensity of the line.}
\tablefoottext{*}{Transitions that provide the identification of the species with S/N$\geq$6 in integrated intensity.}}
\label{tab:propanol}
\end{table*}

\section{Observations}

We have carried out a spectral line survey at 1$\,$mm, 2$\,$mm, 3$\,$mm and 7$\,$mm toward G+0.693 with the Instituto de Radioastronom\'{\i}a Milim\'etrica (IRAM) 30$\,$m telescope (Granada, Spain) and the Yebes 40m telescope (Guadalajara, Spain). The spectral coverage goes from 32$\,$GHz to 233$\,$GHz and the data were smoothed to a typical velocity resolution of $\sim$1.5-2.5$\,$km$\,$s$^{-1}$. The equatorial coordinates of the source are $\alpha$(J2000)=17$^{h}$47$^m$22$^s$ and 
$\delta$(J2000)=$-$28$^\circ$21$'$27$''$. We used the position switching mode with an off position located at (-885$''$, 290$''$) with respect to G+0.693. Each frequency setup was repeated shifting the central frequency by 20-100$\,$MHz in order to identify spurious lines or contamination from the image band. 

The IRAM 30m observations were performed during three sessions in 2019: 10-16th of April, 13-19th of August and 11-15th of December. The EMIR E090, E150, and E230 receivers were used together with the fast Fourier transform spectrometers (FFTS) with a frequency resolution of 200$\,$kHz (equivalent to a velocity resolution between 0.8$\,$km$\,$s$^{-1}$ and 0.26$\,$km$\,$s$^{-1}$ at 3$\,$mm and 1$\,$mm, respectively). The IRAM 30m beam size ranged from $\sim$34$"$ to 11$"$ when going from 72 to 233$\,$GHz. The observations with the Yebes 40m radiotelescope were carried out between the 3rd-9th and 15th-22nd of February 2020. The Nanocosmos Q-band (7 mm) HEMT receiver was used which enabled ultra broad-band observations with an instantaneous bandwidth of 18.5 GHz per polarization between 31.3$\,$GHz and 50.6$\,$GHz \citep[][]{tercero21}. The 16 FFTS provided a channel width of 38$\,$kHz, which corresponds to a velocity resolution of 0.22-0.36$\,$km$\,$s$^{-1}$. The Yebes 40m beam size ranged from $\sim$54$"$ to 36$"$ when moving from 31 to 50$\,$GHz. The line intensity of all spectra was measured in units of T$_A$$^{*}$ as the molecular emission toward G+0.693 is extended compared to the main beam of the telescopes \citep[][]{requena-torres06,zeng20}. 

\section{Analysis and results}

\subsection{Detection of the Ga and {\it Aa} conformers of {\it n}-propanol}
\label{propanol}

\begin{figure*}
   \centering
   \includegraphics[angle=0,width=0.8\textwidth]{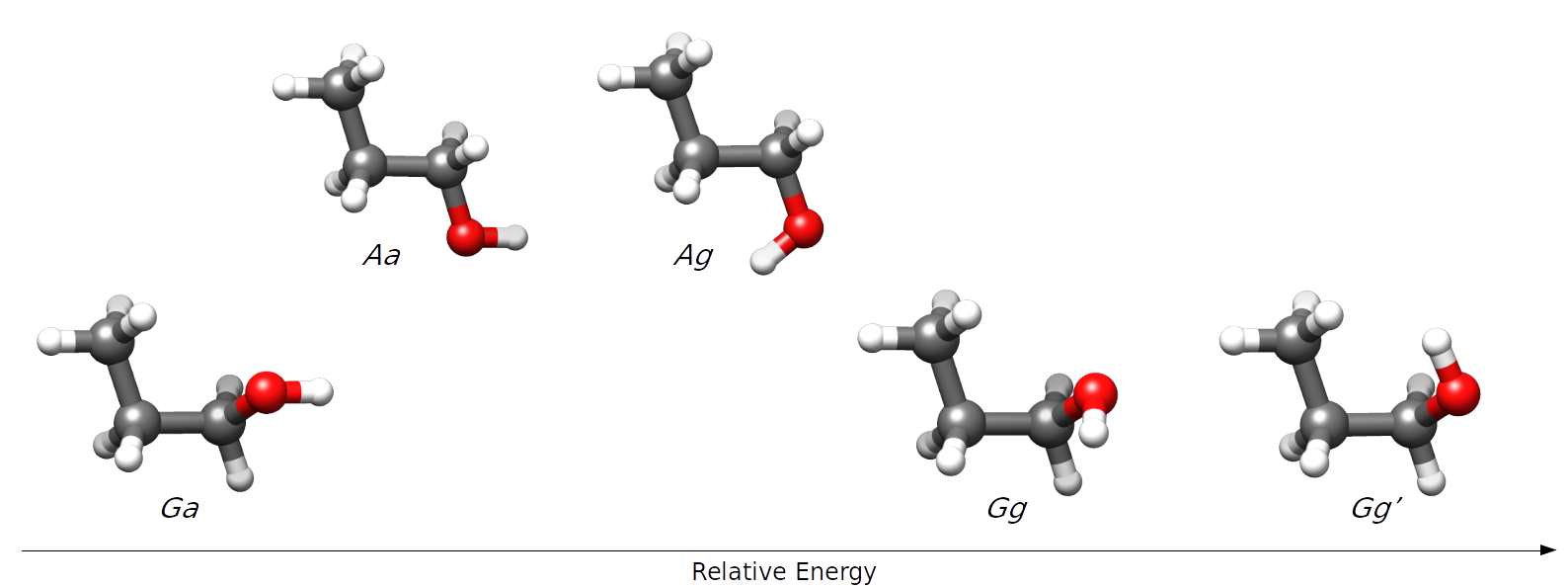}
      \caption{Conformers of $n$-propanol ($n$-C$_3$H$_7$OH) represented as a function of relative energy. The two lowest energy conformers are the {\it Ga} and {\it Aa} conformers, which have been detected toward G+0.693 (see Section$\,$\ref{propanol}). Note that the x-axis is not to scale for visualisation purposes. The relative energies of the different conformers of $n$-propanol are: E=0 K for {\it Ga}, E=40.3 K for {\it Aa}, E=64.7 K for {\it Ag}, E=68.8 K for {\it Gg}, and E=73.2 K for {\it Gg$'$} (see also Section$\,$\ref{propanol}).}
      \label{fig0}
   \end{figure*}

We have used the data analysis package MADCUBA\footnote{MADCUBA, or MAdrid Data CUBe Analysis, is a software developed at the Center of Astrobiology in Madrid: https://cab.inta-csic.es/madcuba/index.html} \citep[][]{martin19} to search for all transitions from the {\it Ga}, {\it Aa}, {\it Gg}, and {\it Gg$'$} conformers of {\it n}-propanol (or 1-propanol) within our dataset (see Figure$\,$\ref{fig0}). We could not search for the {\it Ag} conformer because its energy levels are heavily perturbed by the $Aa$ species and the spectroscopic parameters have never been reported \citep[see][]{kisiel10}. The relative energies of the {\it n}-propanol conformers are: E=0 K for {\it Ga}, E=40.3 K for {\it Aa}, E=64.7 K for {\it Ag}, E=68.8 K for {\it Gg}, and E=73.2 K for {\it Gg$'$} \citep[][]{kisiel10}. 
We note, however, that the conformers of $n$-propanol were modelled independently (i.e. without taking into account their zero-point energy) as it has been done in the past for other species toward G+0.693 such as imines, for which both the low- and high-energy conformers have been detected \citep[see][]{rivilla19}.
Our data reveal that only the two lowest energy conformers, {\it Ga} and {\it Aa} (hereafter {\it Ga}-{\it n}-C$_3$H$_7$OH and {\it Aa}-{\it n}-C$_3$H$_7$OH, respectively; see Figure$\,$\ref{fig0}), are detected toward G+0.693 (see below). The spectroscopic information of {\it Ga}-{\it n}-C$_3$H$_7$OH has been extracted from the Cologne Database for Molecular Spectroscopy catalog \cite[CDMS, entry 060505;][]{endres16}, which has been derived from the rotational spectrum measured by \citet{kisiel10}. The spectroscopic information for the rest of the conformers of $n$-propanol, has been obtained by us using the data from \citet{dreizler81} and \citet{kisiel10}.

\begin{figure*}
   \centering
   \includegraphics[angle=270,width=1.0\textwidth]{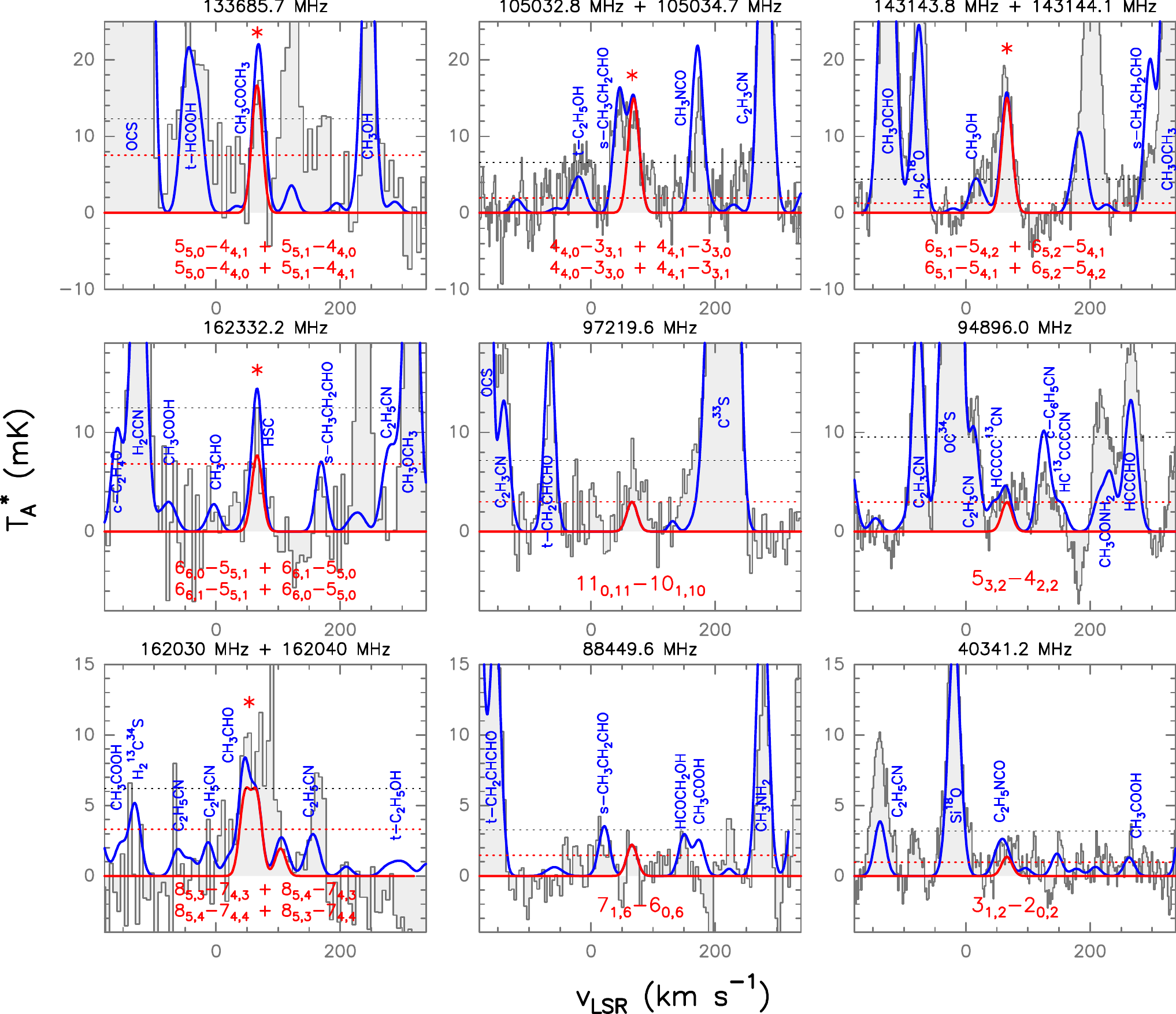}
      \caption{Unblended and slightly blended transitions of {\it Ga}-{\it n}-C$_3$H$_7$OH detected toward G+0.693. Histogram and grey shaded area correspond to the observed IRAM 30m and Yebes 40m spectra. Red lines indicate the best LTE fit to the lines of {\it Ga}-{\it n}-C$_3$H$_7$OH reported in Table$\,$\ref{tab:propanol}. Blue lines show the overall fit to the observed spectra considering all molecular species detected in G+0.693. The quantum numbers of the {\it Ga}-{\it n}-C$_3$H$_7$OH transitions are indicated in red at the bottom part of the panels, while blue labels present the species identified in the shown spectra. The frequencies of the lines are shown in the upper part of each panel. The red asterisks indicate the transitions that provide the identification of {\it Ga}-{\it n}-C$_3$H$_7$OH. Black dotted lines indicate the 3$\times$rms level in peak intensity in the spectra. Red dotted lines show the detectability limit in integrated area defined as 3$\times$rms/$\sqrt{N}$, with $N$ the number of channels across the linewidth.}
      \label{fig:Ga}
   \end{figure*}

The line identification and analysis of the transitions of {\it Ga}-{\it n}-C$_3$H$_7$OH and {\it Aa}-{\it n}-C$_3$H$_7$OH was carried out using the SLIM (Spectral
Line Identification and Modelling) tool of MADCUBA. SLIM uses the spectroscopic data entries from the molecular catalogs to generate synthetic spectra under the assumption of Local Thermodynamic Equilibrium (LTE) and by considering line opacity effects. The LTE fit that best matches the observed line spectra of {\it Ga}-{\it n}-C$_3$H$_7$OH and {\it Aa}-{\it n}-C$_3$H$_7$OH provides the derived values for the molecular column densities (N), excitation temperatures (T$_{ex}$), central radial velocities of the gas (v$_{LSR}$) and full widths at half maximum (FWHM) of the line emission. 

\begin{figure*}
   \centering
   \includegraphics[angle=270,width=1.0\textwidth]{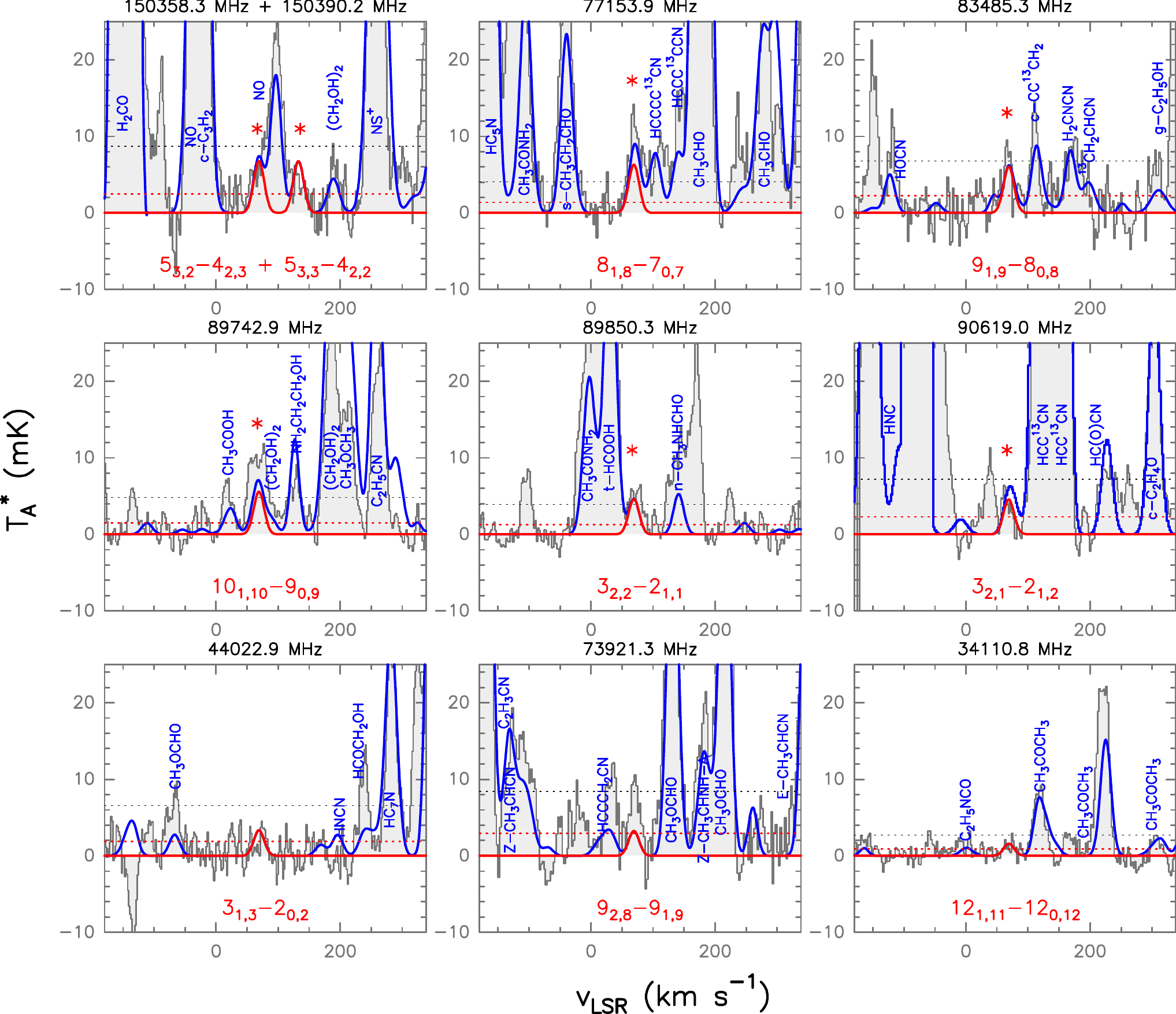}
      \caption{Unblended and slightly blended transitions of {\it Aa}-{\it n}-C$_3$H$_7$OH detected toward G+0.693. Histogram, grey shaded area, color lines and labels are as in Figure$\,$\ref{fig:Ga}. Red lines indicate the best LTE fit to the lines of {\it Aa}-{\it n}-C$_3$H$_7$OH reported in Table$\,$\ref{tab:propanol}. The frequencies of the lines are shown in the upper part of each panel, while the red asterisks indicate the transitions that contribute to the identification of {\it Aa}-{\it n}-C$_3$H$_7$OH. Black dotted lines indicate the 3$\times$rms level in peak intensity in the spectra. Red dotted lines show the detectability limit in integrated area defined as 3$\times$rms/$\sqrt{N}$, with $N$ the number of channels across the linewidth.}
      \label{fig:Aa}
   \end{figure*}

Figures$\,$\ref{fig:Ga} and \ref{fig:Aa} present the unblended and slightly blended transitions of {\it Ga}-{\it n}-C$_3$H$_7$OH and {\it Aa}-{\it n}-C$_3$H$_7$OH, respectively, detected toward G+0.693. Table$\,$\ref{tab:propanol} reports the spectroscopic information of the transitions from these species together with the rms noise level measured in the spectra (rms), the derived integrated line intensities ($\int T_A^* d\nu$) and the signal-to-noise (S/N) ratios (also in integrated intensity) obtained with MADCUBA-SLIM. The rms noise level (rms) was calculated at the velocity resolution reported in Table$\,$\ref{tab:propanol} using all the line-free parts of each spectrum within a velocity range of $\pm$200$\,$km$\,$s$^{-1}$ around each line. For those spectra where not enough line-free parts were available, a velocity range of $\pm$600$\,$km$\,$s$^{-1}$ around each line was used. To improve the visualization of the detected lines, some of the spectra have been smoothed to a velocity resolution of 4-8$\,$km$\,$s$^{-1}$, enough to provide $\geq$3 resolution elements over the linewidth. We stress that the smoothing of the lines does not affect our analysis in any form as demonstrated by \citet[][]{rivilla20}. 

In Figures $\,$\ref{fig:Ga} and \ref{fig:Aa} and Table$\,$\ref{tab:propanol}, we indicate with asterisks the transitions that provide the identification of {\it Ga}-{\it n}-C$_3$H$_7$OH and {\it Aa}-{\it n}-C$_3$H$_7$OH. All these transitions have S/N$\geq$6, where the error in the integrated intensity of the lines is calculated as $\rm rms$$\times$$\sqrt{\delta \rm v \times \rm \Delta v}$, where $\delta \rm v$ is the velocity resolution of the spectra and $\rm \Delta v$ is the FWHM of the line emission. Note, however, that if a more conservative approach were taken (i.e. if we considered $\rm \Delta v$ as the full velocity range covered by the line), then the S/N ratios would be lower by a factor sqrt(2)=1.4. This would not affect the identification of the species reported here because all the bright lines that contribute to the identification of {\it Ga}-{\it n}-C$_3$H$_7$OH and {\it Aa}-{\it n}-C$_3$H$_7$OH would have S/N$\geq$4.5 even after correcting by this factor. The rest of transitions of {\it Ga}-{\it n}-C$_3$H$_7$OH and {\it Aa}-{\it n}-C$_3$H$_7$OH reported in Table$\,$\ref{tab:propanol} and Figures $\,$\ref{fig:Ga} and \ref{fig:Aa} without asterisks, demonstrate that they are consistent with the species identification and their LTE fits. Figures$\,$\ref{fig:Ga} and \ref{fig:Aa} also show the 3$\times$rms noise level in peak intensity (black dotted lines) and the 3$\times$rms/$\sqrt{N}$ level (with $N$ the number of channels across the FWHM), which is associated with the noise level in integrated intensity, i.e. our detectability criterion (see red dotted lines).

The 6$_{6,X}$$\rightarrow$5$_{5,X}$, 4$_{4,X}$$\rightarrow$3$_{3,X}$, and 3$_{1,2}$$\rightarrow$2$_{0,2}$ transitions of {\it Ga}-{\it n}-C$_3$H$_7$OH at 162332$\,$MHz, 105032$\,$MHz, and 40341$\,$MHz, respectively (see Figure$\,$\ref{fig:Ga}), are partly blended with HSC, $s$-propanal and C$_2$H$_5$NCO lines of similar intensity \citep[see the detections of some of these species in][]{rivilla20,rodriguez-almeida21b}. However, we note that the global LTE fit (i.e. considering all molecular species) of the spectra at these frequencies, matches perfectly the observed spectra (see black histograms and blue solid lines in Figure$\,$\ref{fig:Ga}). For {\it Aa}-{\it n}-C$_3$H$_7$OH (see Figure$\,$\ref{fig:Aa}), the 10$_{1,10}$$\rightarrow$9$_{0,9}$ and 9$_{2,8}$$\rightarrow$9$_{1,9}$ transitions are likely blended with lines from unidentified species, besides a small contribution from (CH$_2$OH)$_2$ for the 10$_{1,10}$$\rightarrow$9$_{0,9}$ line (see Table$\,$\ref{tab:propanol}). 
The derived S/N ratio of the 3$_{1,3}$$\rightarrow$2$_{0,2}$ transition should be taken with caution since the LTE fit performed by MADCUBA-SLIM overpredicts the integrated intensity of this line.
The rest of {\it Ga}-{\it n}-C$_3$H$_7$OH and {\it Aa}-{\it n}-C$_3$H$_7$OH transitions considered in our analysis but not shown here, are consistent with their MADCUBA-SLIM LTE fits but appear heavily blended with lines from other molecular species. 

The best LTE fit of the {\it Ga}-{\it n}-C$_3$H$_7$OH lines is obtained for a column density of N({\it Ga}-{\it n}-C$_3$H$_7$OH)=(5.5$\pm$0.4)$\times$10$^{13}$$\,$cm$^{-2}$ and a T$_{ex}$=14$\,$K, V$_{LSR}$=65.9$\pm$1.0$\,$km$\,$s$^{-1}$ and $\Delta v$=20$\,$km$\,$s$^{-1}$ (T$_{ex}$ and $\Delta v$ needed to be fixed for the SLIM-AUTOFIT algorithm  to find convergence). For {\it Aa}-{\it n}-C$_3$H$_7$OH, the best fit of the lines was reached for a N({\it Aa}-{\it n}-C$_3$H$_7$OH)=(3.4$\pm$0.3)$\times$10$^{13}$$\,$cm$^{-2}$ and a T$_{ex}$=12.5$\pm$2.0$\,$K, after fixing the central radial velocity and the FWHM of the source to  V$_{LSR}$=69$\,$km$\,$s$^{-1}$ and $\Delta v$=20$\,$km$\,$s$^{-1}$. The derived and assumed T$_{ex}$, V$_{LSR}$ and $\Delta v$ are consistent with those measured for other molecular species toward this source \citep[see e.g.][]{requena-torres06,zeng18,rivilla20,rodriguez-almeida21a,rivilla21a}, and with the results obtained from the rotational diagrams of {\it Ga}-{\it n}-C$_3$H$_7$OH and {\it Aa}-{\it n}-C$_3$H$_7$OH shown in Figures$\,$\ref{fig:rotdiag-Ga} and \ref{fig:rotdiag-Aa} in the Appendix.

To derive the abundances of the {\it Ga} and {\it Aa} conformers of propanol, we assume a H$_2$ column density of 1.35$\times$10$^{23}$$\,$cm$^{-2}$ as measured toward G+0.693 \citep[][]{martin08}. This provides an abundance of (4.1$\pm$0.3)$\times$10$^{-10}$ for {\it Ga}-{\it n}-C$_3$H$_7$OH and of (2.5$\pm$0.2)$\times$10$^{-10}$ for {\it Aa}-{\it n}-C$_3$H$_7$OH. The abundance ratio between the two detected conformers of {\it n}-propanol is {\it Ga}/{\it Aa}=1.64. Considering this ratio and that the {\it Aa} conformer lies $\sim$40$\,$K above {\it Ga}, one can check whether these two isomers are in thermodynamic equilibrium. By using Equation$\,$3 of \citet{rivilla19}, the derived T$_{kin}$ is $\sim$80$\,$K, consistent with the kinetic temperature range measured from CH$_3$CN toward G+0.693 \citep[between 70 and 150$\,$K; see][]{zeng18}. The calculated {\it Ga}/{\it Aa} abundance ratio therefore suggests that both species are in thermodynamic equilibrium (understood as the equilibration between the abundances of the {\it Ga} and {\it Aa} conformers) as observed for other species such as the E/Z isomers of imines \citep[see][]{garcia21}.

\begin{table*}
\caption{Upper limits obtained for the Gg and Gg$'$ conformers of {\it n}-propanol, for the $anti$ and $gauche$ conformers of i-propanol, and for the $trans$ conformer of ethyl methyl ether.}
\scriptsize
\centering
\begin{tabular}{ccccccccc}
\hline\hline
Species & Formula & Frequency & QNs
& E$_{\mathrm{u}}$ &$\log\mathrm{I(300K)}$ & rms & N\tablefootmark{a} & Abundance\tablefootmark{b} \\
& & (MHz) & (J$''_{\mathrm{K_a'',K_c''}}$- J$'_{\mathrm{K_a',K_c'}}$) & (K) & (nm$^{2}$$\,$MHz) & (mK) & ($\mathrm{cm^{-2}}$) & \\ \hline
{\it n}-propanol ($Gg$) & $Gg$-{\it n}-C$_3$H$_7$OH & 75699.1096 & $3_{3,1}-2_{2,1}$ & 75.7 & -6.1471  & 2.3 & $\leq$1.6$\times$10$^{13}$\tablefootmark{c} & $\leq$1.2$\times$10$^{-10}$ \\
& & 75700.5781 & $3_{3,0}-2_{2,1}$ & 75.7 & -6.2273  & 2.3 & - & - \\
{\it n}-propanol ($Gg'$) & $Gg'$-{\it n}-C$_3$H$_7$OH & 87730.9823 & $10_{1,10}-9_{1,9}$ & 97.0 & -5.2641  & 0.9 & $\leq$5.7$\times$10$^{12}$ & $\leq$4.2$\times$10$^{-11}$ \\ \hline 
isopropanol ({\it gauche}) & $g$-$i$-C$_3$H$_7$OH & 88532.3081 & $4_{1,4}-3_{1,3}$ & 7.5 & -5.7726  & 1.8 & $\leq$8.5$\times$10$^{12}$\tablefootmark{c} & $\leq$6.3$\times$10$^{-11}$ \\ 
& & 88538.5001 & $4_{0,4}-3_{0,3}$ & 7.5 & -5.7737  & 1.8 & - & - \\
isopropanol ({\it anti}) & $a$-$i$-C$_3$H$_7$OH & 79734.1627 & $8_{1,8}-7_{0,7}$ & 137.9 & -5.5140  & 2.7 & $\leq$3.9$\times$10$^{12}$\tablefootmark{c} & $\leq$2.9$\times$10$^{-11}$ \\ 
& & 79734.1627 & $8_{0,8}-7_{1,7}$ & 137.9 & -5.5140  & 2.7 & - & - \\
\hline
ethyl methyl ether ({\it trans}) & $t$-CH$_3$OCH$_2$CH$_3$ & 73243.2848 & $4_{2,3}-4_{1,4}$ & 8.5 & -5.6357  & 2.7 & $\leq$3.8$\times$10$^{13}$ & $\leq$2.8$\times$10$^{-10}$ \\
\hline
\label{tab:conformers}
\end{tabular}
\tablefoot{
\tablefoottext{a}{Upper limits to the column densities calculated assuming a T$_{ex}$=12.5 K, i.e. the same as the one derived for {\it Aa}-{\it n}-C$_3$H$_7$OH, and after fixing V$_{LSR}$=69 km$\,$s$^{-1}$ and $\Delta$v=20 km$\,$s$^{-1}$ (see Section$\,$\ref{propanol}).}
\tablefoottext{b}{Abundances were derived considering a H$_2$ column density of 1.35$\times$10$^{23}$$\,$cm$^{-2}$ \citep[][]{martin08}.}
\tablefoottext{c}{Upper limits were obtained considering the two transitions shown in the Table, which appear blended in the spectra. The selected transitions are the cleanest and the brightest of all the lines covered in our observations.}
}
\end{table*}

The derived upper limits to the other two conformers of $n$-propanol for which spectroscopic data exist ({\it Gg} and {\it Gg$'$}), are presented in Table$\,$\ref{tab:conformers}. They have been calculated using the rms noise level of the brightest, less-blended transition spectra, and assuming a T$_{ex}$=12.5$\,$K, i.e. the same T$_{ex}$ as the one inferred from {\it Aa}-{\it n}-C$_3$H$_7$OH (see above). The derived upper limits to their column densities and abundances are, respectively, $\leq$1.6$\times$10$^{13}$$\,$cm$^{-2}$ and $\leq$1.2$\times$10$^{-10}$ for $Gg$-$n$-C$_3$H$_7$OH, and $\leq$5.7$\times$10$^{12}$$\,$cm$^{-2}$ and $\leq$4.2$\times$10$^{-11}$ for $Gg'$-$n$-C$_3$H$_7$OH (see Table$\,$\ref{tab:conformers}). 

By using the abundance ratios {\it Ga}/{\it Gg} and {\it Ga}/{\it Gg$'$}, we can  investigate whether the {\it Gg} and {\it Gg$'$} conformers are also in thermodynamic equilibrium. The derived abundance ratios {\it Ga}/{\it Gg}$\geq$3.4 and {\it Ga}/{\it Gg$'$}$\geq$9.7 yield kinetic temperatures T$_{kin}$$\leq$56$\,$K and T$_{kin}$$\leq$32$\,$K, respectively. These kinetic temperatures are lower than the ones measured toward G+0.693 \citep[][]{zeng18}, which suggests that only the lower energy conformers of a molecular species could reach thermodynamic equilibrium via mechanisms such as multi-dimensional quantum tunneling as found for imines and acids \citep[see][]{garcia21,garcia22}.

\subsection{Search of isopropanol (i-propanol) and ethyl methyl ether}

Isopropanol (also named isopropyl alcohol and 2-propanol) is a structural isomer of {\it n}-propanol. Its chemical formula is CH$_3$CH(OH)CH$_3$. We have searched for the {\it gauche} and {\it anti} conformers of {\it i}-propanol ({\it g}-{\it i}-C$_3$H$_7$OH and {\it a}-{\it i}-C$_3$H$_7$OH, respectively) in our dataset by using entries 060518 and 060519 of the CDMS molecular catalog \citep[][]{endres16}, obtained using the spectroscopic work of \citet{maeda06}. The {\it anti} conformer is about 120$\,$K higher in energy than the doubly-degenerate {\it gauche} conformer. 

None of the two conformers of {\it i}-propanol are detected (see Table$\,$\ref{tab:conformers}), with upper limits to their column densities of $\leq$8.5$\times$10$^{12}$$\,$cm$^{-2}$ for {\it g}-{\it i}-C$_3$H$_7$OH and $\leq$3.9$\times$10$^{12}$$\,$cm$^{-2}$ for {\it a}-{\it i}-C$_3$H$_7$OH. These column densities translate into abundance upper limits of $\leq$6.3$\times$10$^{-11}$ and $\leq$2.9$\times$10$^{-11}$ for the {\it gauche} and {\it anti} conformers of {\it i}-propanol, respectively.

Propanol has a third structural isomer: ethyl methyl ether (CH$_3$OCH$_2$CH$_3$ or methoxyethane). We have also searched for the $trans$ conformer of this species (the lowest in energy) in our dataset by using the spectroscopic information reported by \citet{fuchs03} and by inserting the entry for this species in CDMS/JPL format into MADCUBA. This species has not been detected toward G+0.693 either, with upper limits to its column density and abundance of, respectively, $\leq$3.8$\times$10$^{13}$$\,$cm$^{-2}$ and $\leq$2.8$\times$10$^{-10}$ (see Table$\,$\ref{tab:conformers}).

\subsection{Other alcohols: detection of the {\it syn} and {\it anti} forms of vinyl alcohol}
\label{others}

We have also searched for other alcohols within our spectroscopic survey toward G+0.693. We have targeted vinyl alcohol (or ethenol, with formula H$_2$C=CHOH), cyclopropanol (c-C$_3$H$_5$OH) and {\it n}-butanol ({\it n}-C$_4$H$_9$OH).
We have used the CDMS entries 044506 and 044507 for the {\it syn} and {\it anti} forms of vinyl alcohol \citep[{\it s}-H$_2$C=CHOH and {\it a}-H$_2$C=CHOH in CDMS;][]{endres16} based on the spectroscopic work of \citet{melosso19}. The {\it syn} form of vinyl alcohol is the lowest in energy with the {\it anti} form lying about 4.6 kJ/mol (or 550$\,$K) above the {\it syn} form \citep{rodler85}. For cyclopropanol (c-C$_3$H$_5$OH), we have taken the spectroscopic information for the {\it gauche} conformer (the most stable one) from the work of \citet{macdonald78} and inserted the entry for this species in CDMS/JPL format into MADCUBA. In the same way, for {\it n}-butanol (both {\it AGa}-(E) and {\it AGa}-(A) methyl internal rotation substates), we have calculated their spectroscopic information from the laboratory experiments and theoretical calculations of \citet{kawashima21}.

\begin{figure*}
   \centering
   \includegraphics[angle=270,width=1.0\textwidth]{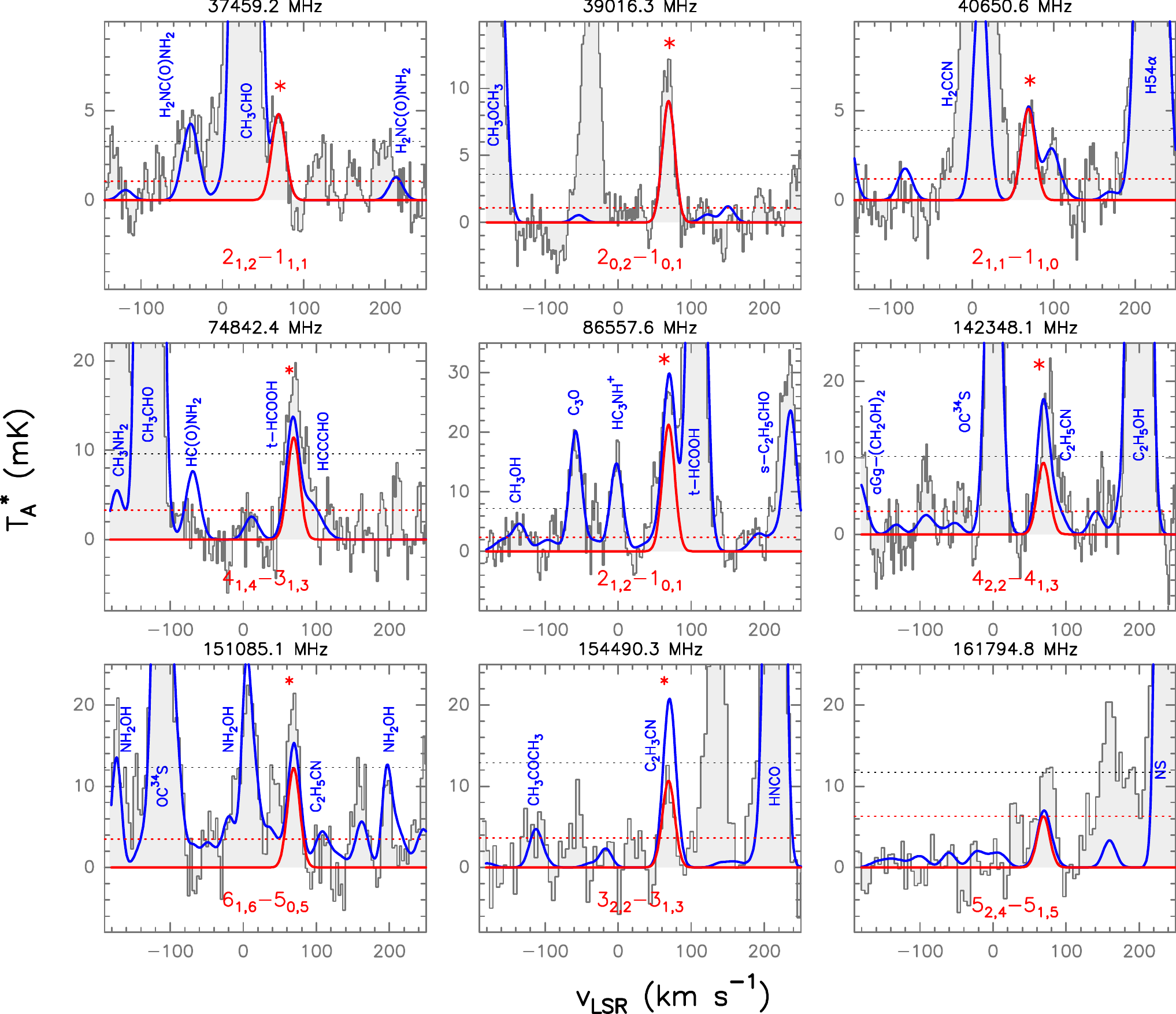}
      \caption{\label{fig:s-vinylalcohol} Unblended and slightly blended transitions of  {\it s}-H$_2$C=CHOH detected toward G+0.693. For a description of the colours displayed in the plot, see Figure~\ref{fig:Ga}. The spectroscopic information and the derived parameters from the best LTE fit of these lines, are reported in Table~\ref{tab:vinylalcohol}. The frequencies of the lines are shown in the upper part of each panel, while the red asterisks indicate the transitions that contribute to the identification of {\it s}-H$_2$C=CHOH. Black dotted lines indicate the 3$\times$rms level in peak intensity in the spectra. Red dotted lines show the detectability limit in integrated area defined as 3$\times$rms/$\sqrt{N}$, with $N$ the number of channels across the linewidth.}
   \end{figure*}

\begin{figure*}
   \centering
   \includegraphics[angle=270,width=1.0\textwidth]{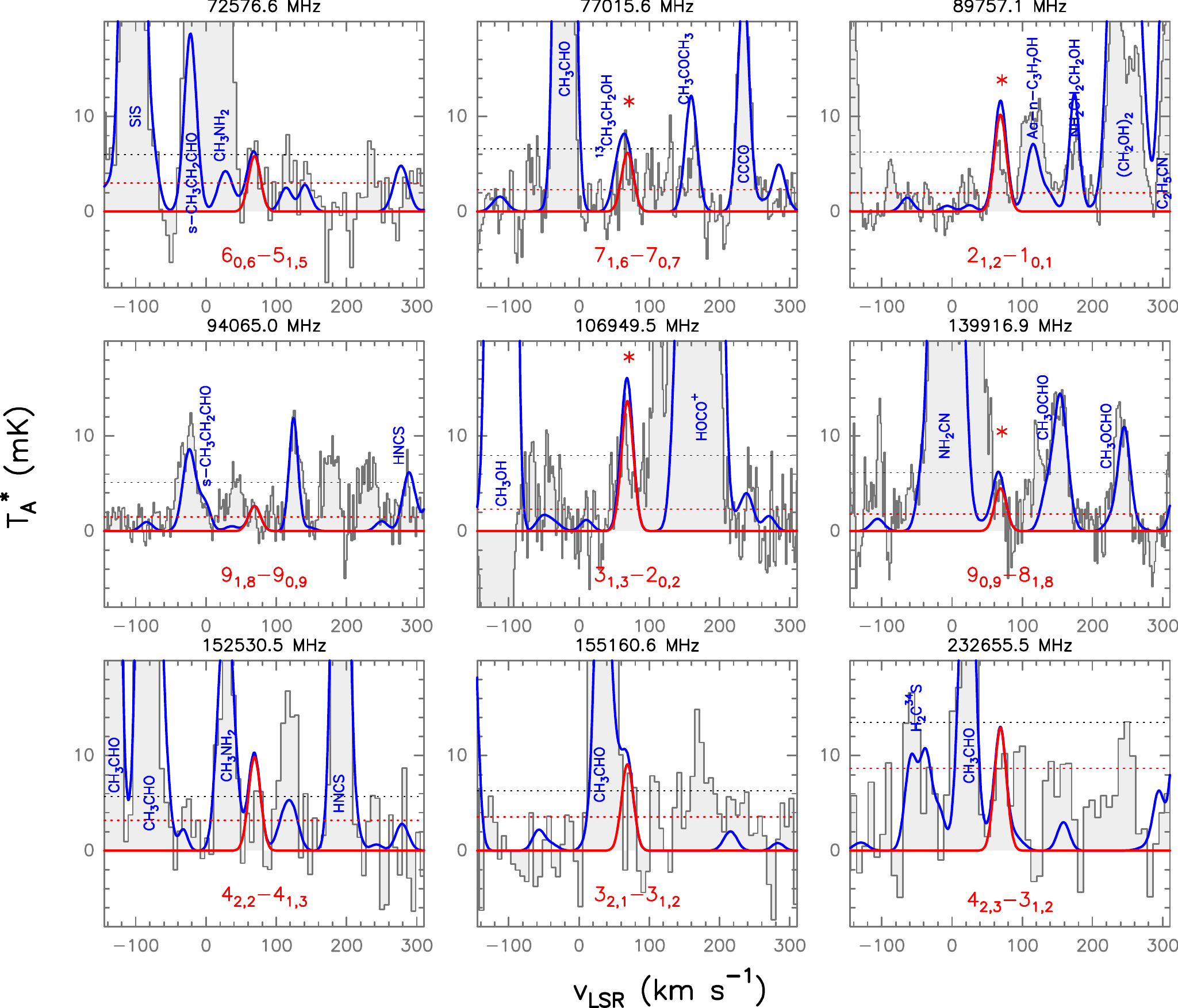}
      \caption{\label{fig:a-vinylalcohol} Unblended and slightly blended transitions of  {\it a}-H$_2$C=CHOH detected toward G+0.693. For a description of the colours displayed in the plot, see Figure~\ref{fig:Ga}. The spectroscopic information and the derived parameters from the best LTE fit of these lines, are reported in Table~\ref{tab:vinylalcohol}. The frequencies of the lines are shown in the upper part of each panel. Red asterisks indicate the transitions that contribute to the identification of {\it a}-H$_2$C=CHOH. Black dotted lines indicate the 3$\times$rms level in peak intensity in the spectra. Red dotted lines show the detectability limit in integrated area defined as 3$\times$rms/$\sqrt{N}$, with $N$ the number of channels across the linewidth.}
   \end{figure*}

Of all these species, only vinyl alcohol has been detected in both {\it syn} and {\it anti} forms ( {\it s}-H$_2$C=CHOH and  {\it a}-H$_2$C=CHOH, respectively) toward G+0.693. In Table$\,$\ref{tab:vinylalcohol} and Figures$\,$\ref{fig:s-vinylalcohol} and \ref{fig:a-vinylalcohol}, we report the clean and slightly blended transitions of the {\it syn} and {\it anti} forms of H$_2$C=CHOH detected toward G+0.693 with a S/N$\geq$6 (see the transitions with asterisks). The rest of lines without asterisks are also shown in Table$\,$\ref{tab:vinylalcohol} and Figures$\,$\ref{fig:s-vinylalcohol} and \ref{fig:a-vinylalcohol}, to demonstrate that they are consistent with the identification of these species and their LTE fits. 

Table$\,$\ref{tab:vinylalcohol} presents 4 clean transitions and 3 slightly blended transitions of  {\it s}-H$_2$C=CHOH, and 3 clean transitions and 1 slightly blended transitions of  {\it a}-H$_2$C=CHOH. The 4$_{2,2}$$\rightarrow$4$_{1,3}$ and 3$_{2,2}$$\rightarrow$3$_{1,3}$ transitions of $s$-H$_2$C=CHOH are blended with lines of C$_2$H$_5$CN and C$_2$H$_3$CN of similar intensity respectively, while the 3$_{2,1}$$\rightarrow$3$_{1,2}$ transition of {\it a}-H$_2$C=CHOH is blended with CH$_3$CHO (see Table$\,$\ref{tab:vinylalcohol}). 
The 4$_{2,2}$$\rightarrow$4$_{1,3}$ and 3$_{2,1}$$\rightarrow$3$_{1,2}$ transitions of $a$-H$_2$C=CHOH also present S/N$\geq$6. However, these ratios are overestimated and should be taken with caution since they are calculated from the LTE fit of MADCUBA-SLIM and not from the gaussian fit of the individual lines (see Table$\,$\ref{tab:vinylalcohol}). As a result, these lines are not considered for the identification of {\it a}-H$_2$C=CHOH. 

We have derived the physical parameters of  {\it s}-H$_2$C=CHOH and  {\it a}-H$_2$C=CHOH using the SLIM tool from MADCUBA \citep[][]{martin19}. The best LTE fit for  {\it s}-H$_2$C=CHOH is obtained for a column density of (1.11$\pm$0.08)$\times$10$^{14}$$\,$cm$^{-2}$ and an excitation temperature of T$_{ex}$=8.0$\pm$1.0$\,$K. For  {\it a}-H$_2$C=CHOH, the best fit of the lines is obtained for a column density of (1.3$\pm$0.4)$\times$10$^{13}$$\,$cm$^{-2}$ after fixing T$_{ex}$=14$\,$K (as we did for {\it Ga}-{\it n}-propanol), $\Delta \rm v$=20$\,$km$\,$s$^{-1}$ and V$_{LSR}$=69$\,$km$\,$s$^{-1}$. These results are consistent with those obtained through the rotational diagrams of {\it s}-H$_2$C=CHOH and  {\it a}-H$_2$C=CHOH (see Figures$\,$\ref{fig:rotdiag-s} and \ref{fig:rotdiag-a} in the Appendix). Assuming a H$_2$ column density of 1.35$\times$10$^{23}$$\,$cm$^{-2}$ \citep[][see also Section$\,$\ref{propanol}]{martin08}, the derived column densities translate into molecular abundances of (8.2$\pm$0.6)$\times$10$^{-10}$ for  {\it s}-H$_2$C=CHOH, and of (9.6$\pm$3.0)$\times$10$^{-11}$ for  {\it a}-H$_2$C=CHOH. The {\it syn}/{\it anti} abundance ratio for H$_2$C=CHOH is thus $\sim$8.5. 

By using again Equation$\,$3 from \citet{rivilla19}, we find a T$_{kin}$$\sim$260$\,$K, which is slightly higher than that measured toward G+0.693 \citep[][]{zeng18}. Note that if we assumed a T$_{ex}$=8.0$\,$K (i.e. the one obtained for {\it s}-H$_2$C=CHOH), the derived column density would decrease by just a factor of 1.2, which would translate into a $syn$/$anti$ abundance ratio $\sim$10. This ratio would be consistent with a T$_{kin}$$\sim$240$\,$K, which is still higher than that determined by \citet[][]{zeng18}. This may mean that the two isomers of vinyl alcohol are not in thermodynamic equilibrium. 

Finally, in Table$\,$\ref{tab:others}, we report the upper limits to the column densities and abundances obtained for cyclopropanol and n-butanol. The upper limits to their abundance are $\leq$1.4$\times$10$^{-10}$ and $\leq$4.1$\times$10$^{-11}$, respectively. 

\begin{table*}
\caption{Selected lines of both forms of vinyl alcohol (H$_2$C=CHOH), {\it syn (s)} and {\it anti (a)}, detected towards G+0.693.}
\scriptsize
\centering
\begin{tabular}{ccccccccc}
\hline\hline
Frequency & QNs
& E$_{\mathrm{u}}$ &$\log\mathrm{I(300K)}$ &rms& $\delta v$ & $\int T_A^* d\nu$&S/N\tablefootmark{a}&blending \\
(MHz)&(J$''_{\mathrm{K_a'',K_c''}}$- J$'_{\mathrm{K_a',K_c'}}$)& (K) & (nm$^{2}$$\,$MHz) &(mK)& ($\mathrm{km\,s^{-1}}$) & ($\mathrm{K\,km\,s^{-1}}$)& & \\
\hline
\multicolumn{9}{c}{{\bf  {\it s}-H$_2$C=CHOH}} \\ \hline
37459.19\tablefootmark{*} & $2_{1,2}-1_{1,1}$ & 5.09  & -6.3478 & 1.1 & 2.0 & 0.103(0.007)\tablefootmark{c} & 14.7 & clean transition\tablefootmark{b}\\
39016.31\tablefootmark{*} & $2_{0,2}-1_{0,1}$ & 2.81  & -6.1842 & 1.2 & 1.9 &0.193(0.007) & 25.9 & clean transition\tablefootmark{b}\\
40650.65\tablefootmark{*} & $2_{1,1}-1_{1,0}$ & 5.32  & -6.2770 & 1.3 & 1.9 &0.109(0.008) & 13.8 & clean transition\tablefootmark{b} \\
74842.40\tablefootmark{*}  & $4_{1,4}-3_{1,3}$ & 11.38 & -5.3566 & 3.2 & 2.4 & 0.244(0.022) & 10.9 & slightly blended with t-HCOOH and HCCHO\tablefootmark{b}\\
86557.56\tablefootmark{*} & $2_{1,2}-1_{0,1}$ & 5.09  & -5.3840 & 2.4 & 2.1 & 0.454(0.016) & 29.1 & slightly blended with t-HCOOH and c-HCOOH\tablefootmark{b}\\
142348.06\tablefootmark{*} & $4_{2,2}-4_{1,3} $& 18.98 & -4.7860 & 3.4 & 1.7 &0.199(0.020) & 10.0 & blended with C$_2$H$_5$CN\\
151085.05\tablefootmark{*} & $6_{1,6}-5_{0,5}$ & 21.24 & -4.5207 & 4.1 & 1.6 & 0.261(0.023) & 11.2 & slightly blended with C$_2$H$_5$CHO\tablefootmark{b}\\
154490.28\tablefootmark{*} & $3_{2,2}-3_{1,3}$ & 15.20 & -4.9105   & 4.3 & 1.6 &0.227(0.024) & 9.4 & blended with C$_2$H$_3$CN \\
161794.81 & $5_{2,4}-5_{1,5}$ & 23.63 & -4.6522  & 3.1 & 6.0 & 0.13(0.04) & 3.1 & clean transition\\ \hline
\multicolumn{9}{c}{{\bf  {\it a}-H$_2$C=CHOH}} \\ \hline
72576.63  & $6_{0,6}-5_{1,5}$ & 19.5 & -4.6023 & 2.0 & 5.0 & 0.096(0.021) & 4.7 & clean transition\\ 
77015.59\tablefootmark{*}  & $7_{1,6}-7_{0,7}$ & 29.6 & -4.2548 & 2.2 & 2.4 & 0.101(0.015) & 6.6 & slightly blended with a-$^{13}$CH$_3$CH$_2$OH\tablefootmark{b} \\ 
89757.10\tablefootmark{*}   & $2_{1,2}-1_{0,1}$ & 5.2 & -4.6954 & 2.1 & 2.0 & 0.167(0.013) & 12.6 & clean transition\tablefootmark{b}\\ 
94064.95  & $9_{1,8}-9_{0,9}$ & 46.0 & -4.0615 & 1.7 & 1.9 & 0.043(0.011) & 4.0 & clean transition \\ 
106949.48\tablefootmark{*} & $3_{1,3}-2_{0,2}$ & 7.9 & -4.4184 & 2.7 & 1.7 & 0.224(0.016) & 14.5 & clean transition\tablefootmark{b} \\ 
139916.89\tablefootmark{*} & $9_{0,9}-8_{1,8}$ & 41.5 & -3.8023 & 2.1 & 1.7 & 0.074(0.012) & 6.1 & clean transition \\
152530.51 & $4_{2,2}-4_{1,3}$ & 19.6 & -4.0754 & 1.9 & 6.4 & 0.162(0.022) & 7.5\tablefootmark{d} & clean transition\tablefootmark{b}\\ 
155160.66 & $3_{2,1}-3_{1,2}$ & 15.8 & -4.2142 & 2.1 & 6.4 & 0.149(0.024) & 6.3\tablefootmark{d} & blended with CH$_3$CHO\tablefootmark{b} \\ 
232655.55 & $4_{2,3}-3_{1,2}$ & 19.5 & -3.7853 & 4.5 & 8.3 & 0.21(0.06) & 3.7 & clean transition\\ \hline 
\end{tabular}
\tablefoot{
\tablefoottext{a}{The S/N ratios are calculated from the integrated intensity of the lines inferred from the LTE fits of MADCUBA-SLIM. Note that they are not calculated from the gaussian fits of individual lines.}
\tablefoottext{b}{These transitions have been detected at the $\geq$3$\,$rms noise level in peak intensity.}
\tablefoottext{c}{The error in the integrated intensity is calculated as $\rm rms$$\times$$\sqrt{\delta \rm v \times \Delta \rm v}$, with $\delta \rm v$ the velocity resolution of the spectra and $\Delta \rm v$ the FWHM of the line emission of 20$\,$km$\,$s$^{-1}$.}
\tablefoottext{d}{This S/N ratio should be taken with caution since the LTE fit carried out by MADCUBA-SLIM overpredicts the integrated intensity of the line.}
\tablefoottext{*}{Transitions that provide the identification of the species with S/N$\geq$6 in integrated intensity.}}
\label{tab:vinylalcohol}
\end{table*}

\begin{table*}
\caption{Derived upper limits to the column densities of other alcohols observed toward G+0.693.}
\centering
\begin{tabular}{cccccccccc}
\hline\hline
Species & Formula & Frequency & QNs 
& E$_{\mathrm{u}}$ &$\log\mathrm{I(300K)}$ & rms & N\tablefootmark{a} & Abundance\tablefootmark{b}\\
& & (MHz) & (J$''_{\mathrm{K_a'',K_c''}}$- J$'_{\mathrm{K_a',K_c'}}$) & (K) & (nm$^{2}$$\,$MHz) & (mK) & ($\mathrm{cm^{-2}}$) & \\
\hline
Cyclopropanol & c-C$_3$H$_5$OH ({\it gauche}) & 93909.6309 & $3_{3,0}-2_{2,1}$ & 8.3 & -5.5796 & 1.4 & $\leq$1.8$\times$10$^{13}$ & $\leq$1.4$\times$10$^{-10}$ \\ 
n-butanol (E conf.) & {\it AGa}-(E)-{\it n}-C$_4$H$_9$OH & 139389.7073 & $6_{6,1}-5_{5,0}$ & 22.2 & -5.1710  & 2.2 & $\leq$5.5$\times$10$^{12}$\tablefootmark{c} & $\leq$4.1$\times$10$^{-11}$ \\
 & & 139389.7073 & $6_{6,0}-5_{5,1}$ & 22.2 & -5.1710  & 2.2 & - & - \\
n-butanol (A conf.) & {\it AGa}-(A)-{\it n}-C$_4$H$_9$OH & 139389.7685 & $6_{6,1}-5_{5,0}$ & 22.2 & -5.1710  & 2.2 & $\leq$6.2$\times$10$^{12}$\tablefootmark{c} & $\leq$4.6$\times$10$^{-11}$ \\
 & & 139389.7685 & $6_{6,0}-5_{5,1}$ & 22.2 & -5.1710  & 2.2 & - & - \\ \hline
\label{tab:others}
\end{tabular}
\tablefoot{
\tablefoottext{a}{Upper limits to the column densities calculated assuming a T$_{ex}$=12.5 K, i.e. the same as the one derived for {\it Aa}-{\it n}-C$_3$H$_7$OH, and after fixing V$_{LSR}$=69 km$\,$s$^{-1}$ and $\Delta$v=20 km$\,$s$^{-1}$ (see Section$\,$\ref{propanol}).}
\tablefoottext{b}{Abundances were derived considering a H$_2$ column density of 1.35$\times$10$^{23}$$\,$cm$^{-2}$ \citep[][]{martin08}.}
\tablefoottext{c}{Upper limits were obtained considering the two transitions shown in the Table, which appear blended in the spectra.}
}
\end{table*}

\section{Discussion}

\subsection{Comparison of the molecular abundances of propanol and butanol with smaller alcohols}
\label{propanol:comparison}

\begin{figure}
   \centering
   \includegraphics[angle=270,width=0.47\textwidth]{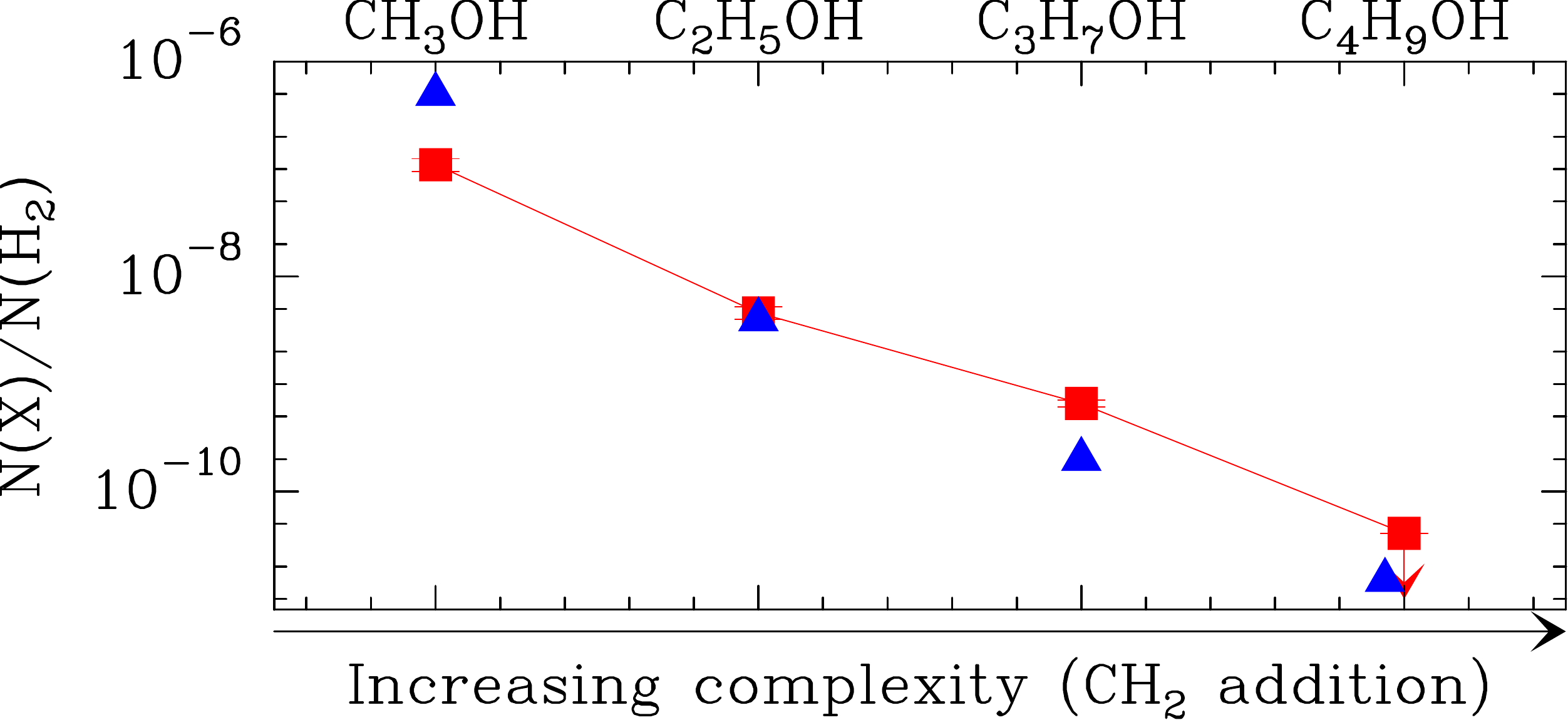}
      \caption{\label{fig:comparison} Comparison of the molecular abundances of the primary alcohols methanol, ethanol, {\it n}-propanol and $n$-butanol measured toward the giant molecular cloud G+0.693 (red squares) with the values predicted by the models of \citet[][blue triangles]{charnley95} for the case with initial solid abundances of 10$^{-6}$, 10$^{-7}$ and 10$^{-8}$ for C$_2$H$_5$OH, propanol and butanol, respectively. Error bars are shown also in red for the measured abundances, and the downward red arrow indicates an upper limit.}
   \end{figure}

In order to gain some insight into the production efficiency of primary alcohols in the ISM, we compare the abundances and upper limits derived for {\it n}-propanol (n-C$_3$H$_7$OH) and n-butanol (n-C$_4$H$_9$OH) with those measured for methanol (CH$_3$OH) and ethanol (C$_2$H$_5$OH) toward G+0.693 (see Figure$\,$\ref{fig:comparison}). The abundances of CH$_3$OH and C$_2$H$_5$OH toward this giant molecular cloud have been provided recently by  \citet{rodriguez-almeida21a} and are (1.1$\pm$0.2)$\times$10$^{-7}$ for CH$_3$OH and (4.6$\pm$0.6)$\times$10$^{-9}$ for C$_2$H$_5$OH. 

From Figure$\,$\ref{fig:comparison}, it is clear that primary alcohols are produced about ten times less each time a carbon atom is added to the chain (see red squares). Indeed, C$_2$H$_5$OH, n-C$_3$H$_7$OH and n-C$_4$H$_9$OH present abundance ratios of, respectively, 0.04, 0.006, and $\leq$0.0004 with respect to CH$_3$OH. Note that the value above and the one shown in Figure$\,$\ref{fig:comparison} for $n$-C$_3$H$_7$OH correspond to the sum of the abundances from both {\it Ga} and {\it Aa} conformers (see Section$\,$\ref{propanol}). The trend of decreasing abundance by roughly one order of magnitude for increasing complexity was already noted by \citet{rodriguez-almeida21a} for CH$_3$OH and C$_2$H$_5$OH as well as for thiols (with the -SH group). Here, we confirm that the same trend stands when increasing the number of carbon atoms to three and four within the chemical structure of primary alcohols. 

By using IRAM 30$\,$m and ALMA data, \citet{tercero15} and \citet{muller16} reported the abundances of CH$_3$OH and C$_2$H$_5$OH and upper limits to the abundance of {\it Ga}-{\it n}-C$_3$H$_7$OH and $i$-propanol toward Orion KL and SgrB2 (N2), respectively. The  C$_2$H$_5$OH/CH$_3$OH abundance ratios are $\sim$0.02 for Orion KL and $\sim$0.05 for SgrB2 (N2), which are consistent with that found toward G+0.693 ($\sim$0.04). The upper limit to the {\it Ga}-{\it n}-C$_3$H$_7$OH/C$_2$H$_5$OH abundance ratio determined toward SgrB2 (N2) is $\leq$0.13, in agreement with the abundance ratio of $\sim$0.14 measured toward G+0.693. For Orion KL, \citet{tercero15} reported a {\it Ga}-{\it n}-C$_3$H$_7$OH/C$_2$H$_5$OH abundance ratio of $\leq$0.017, i.e. a factor of 10 lower than that  obtained toward G+0.693. However, as noted by \citet{muller16}, if the partition functions of C$_2$H$_5$OH and {\it Ga}-{\it n}-C$_3$H$_7$OH are corrected from conformational and vibrational contributions, this ratio becomes $\leq$0.07, which is of the same order of magnitude as the ones found toward SgrB2 (N2) and G+0.693. 


For $i$-propanol, the upper limit to the abundance ratio with respect to CH$_3$OH derived toward G+0.693 is consistent with that measured in the EMoCA survey toward the SgrB2 (N2) massive star-forming region \citep[][]{muller16}. Indeed, while the $i$-propanol abundance upper limit in SgrB2 (N2) provides an $i$-propanol abundance that is $\geq$430 times lower than the abundance of CH$_3$OH, toward G+0.693 our measured upper limit gives an abundance that is $\geq$1700 times lower than that of CH$_3$OH (Table$\,$\ref{tab:conformers}). For the rest of alcohols explored in this work, vinyl alcohol is more than two orders of magnitude less abundant than CH$_3$OH and a factor of 5 less abundant than C$_2$H$_5$OH, while methyl ethyl ether and cyclopropanol have abundances that are, respectively, factors of $\geq$500 and $\geq$800 lower than that of CH$_3$OH (see Tables$\,$\ref{tab:conformers} and \ref{tab:others}). 

Finally, if we compare the derived upper limit abundance of methyl ethyl ether (CH$_3$OCH$_2$CH$_3$) with the abundance of dimethyl ether (CH$_3$OCH$_3$) found toward G+0.693 \citep[of 8.2$\times$10$^{-9}$;][]{requena-torres06}, we find that dimethyl ether is a factor $\geq$30 more abundant than methyl ethyl ether, which is also consistent with the idea that the abundance of these COMs decreases by roughly one order of magnitude with the addition of a $-$CH$_2$ group to their molecular structure.

\subsection{Chemistry of propanol and butanol in G+0.693}
\label{chem:propanol}

The chemistry of large alcohols was theoretically studied by \citet{charnley95} for the case of a massive hot molecular core. In this work, species such as CH$_3$OH, C$_2$H$_5$OH, propanol (both n-C$_3$H$_7$OH and $i$-C$_3$H$_7$OH) and butanol (n-C$_4$H$_9$OH and t-C$_4$H$_9$OH) were assumed to be formed on the surface of dust grains and injected into the gas phase by the thermal evaporation of the grain mantles. The subsequent gas-phase chemistry yields pure ethers through, mainly, alkyl transfer and proton transfer reactions. \citet{charnley95} considered that the initial ice mantle abundance of propanol and butanol were, respectively, factors of 10 and 100 lower than that of solid ethanol. 

In Section$\,$\ref{propanol:comparison}, we reported a similar trend for these large alcohols to decrease their abundance in G+0.693 by approximately a factor of 10 with the addition of a carbon atom. If we thus compare the modelling results of \citet{charnley95} with our measured abundances, the best match is found for the case with initial ice mantle abundances of 10$^{-6}$, 10$^{-7}$ and 10$^{-8}$ for C$_2$H$_5$OH, propanol and butanol, respectively. Indeed, at time-scales of 10$^5$$\,$yrs \citep[the typical time-scales that explain the rich chemistry in complex organics toward Galactic Center giant molecular clouds; see][]{requena-torres06}, the predicted abundance of C$_2$H$_5$OH, propanol and butanol in the gas phase are, respectively, $\sim$4$\times$10$^{-9}$, $\sim$2$\times$10$^{-10}$, and $\sim$1.5$\times$10$^{-11}$ \citep[see Figure$\,$3 in][]{charnley95}. These values are strikingly similar to those measured in G+0.693, i.e. (4.6$\pm$0.6)$\times$10$^{-9}$ for C$_2$H$_5$OH \citep[][]{rodriguez-almeida21a}, (6.6$\pm$0.5)$\times$10$^{-10}$ for propanol, and $\leq$4$\times$10$^{-11}$ for butanol (see blue triangles in Figure$\,$\ref{fig:comparison}). This suggests that, even after gas-phase chemistry takes over, the decrease in the abundance of CH$_3$OH, C$_2$H$_5$OH, propanol and butanol by roughly one order of magnitude with an increasing number of carbon atoms is preserved in the gas phase, which may allow us to indirectly constrain the amount formed in the ices for these alcohols.

We note, however, that while the modelled propanol/C$_2$H$_5$OH abundance ratio is 0.05 \citep[see Figure$\,$3 in][]{charnley95}, the same ratio measured in G+0.693 gives 0.15,  which suggests that propanol is more efficiently produced under the highly energetic processing of the ISM in the Galactic Center than in typical Galactic hot cores (see Section$\,$\ref{propanol:comparison} above). In addition, as pointed out by \citet[][]{requena-torres06}, the models of \citet{charnley95} cannot explain the homogeneity of the COMs abundances across the different GMCs in the Central Molecular Zone. Instead, a continuous replenishment of complex organic material from dust grains into the gas phase by large-scale shocks in the highly turbulent ISM of the Galactic Center, needs to be invoked.  

In the work of \citet{charnley95}, it is assumed that propanol and butanol are present in ices, but it is not explained how these species are formed. In the more recent chemical models of \citet[][]{manigand21}, propanol is formed on the surface of dust grains via two routes: 1) the successive hydrogenation of C$_3$O; and 2) the radical-radical reaction between O and C$_3$H$_8$. The latter has two possible outcomes, O + C$_3$H$_8$$\rightarrow$C$_3$H$_7$OH and O + C$_3$H$_8$$\rightarrow$H$_2$O + C$_3$H$_6$, with branching ratios of 20\% and 80\%, respectively. The models are applied to the case of the IRAS16293-2422 B low-mass hot corino, for which upper limits to the abundance of propanol have been derived \citep[][]{qasim19,manigand21}. The results of \citet{manigand21} only reproduce the factor of 10 difference between the abundances of propanol and C$_2$H$_5$OH in the gas phase at the end of the simulations for models A1-E1, where an efficient destruction of C$_3$ through the gas-phase reaction O + C$_3$$\rightarrow$C$_2$ + CO is considered (see Table$\,$D1 in this work). We note, however, that the absolute abundances of propanol and C$_2$H$_5$OH predicted by these models clearly exceed the values observed toward G+0.693 \citep[of $\geq$10$^{-7}$-10$^{-6}$;][]{manigand21}.

Laboratory experiments carried out by \citet{qasim19} explored alternative formation routes for the formation of propanol in the ices such as the in-situ recombination of HCO (formed after the hydrogenation of CO) with radicals formed via the processing of C$_2$H$_2$. This yields {\it n}-propanol after several hydrogenation steps where propenal (CH$_2$CHCHO) and propanal (CH$_3$CH$_2$CHO) are intermediate steps. These species have been detected toward G+0.693 with abundances of 2.7$\times$10$^{-10}$ for CH$_2$CHCHO and of 1.2$\times$10$^{-9}$ for CH$_3$CH$_2$CHO \citep[see Table$\,$4 in][]{requena-torres08}\footnote{Note that the abundances shown in Table$\,$4 of \citet{requena-torres08} were obtained assuming a H$_2$ column density of 4.1$\times$10$^{22}$$\,$cm$^{-2}$. Therefore, these abudances have been corrected by the H$_2$ column density assumed in this work of 1.35$\times$10$^{23}$$\,$cm$^{-2}$ inferred from C$^{18}$O data by \citet{martin08}.}.  
The abundance of CH$_3$CH$_2$CHO is a factor of $\sim$2 higher than that of {\it n}-propanol, which is consistent with the idea that all these species form concurrently on the surface of dust grains. The non-diffusive mechanism proposed by \citet{qasim19} likely dominates the formation of these species toward G+0.693 since the dust grain temperatures in this source lie below 25$\,$K \citep[][]{rodriguez-fernandez00}.   

\subsection{Vinyl alcohol: comparison with previous observations}
\label{vinyl:comparison}

Vinyl alcohol has been reported tentatively in the past toward the massive hot molecular core SgrB2(N) \citep{turner01}. More recently and thanks to a high-sensitivity spectroscopic survey carried out toward the molecular dark cloud TMC-1 with the Yebes 40m telescope, the {\it syn} conformer of vinyl alcohol ({\it s}-H$_2$C=CHOH) was unambiguously identified \citep{agundez21}. Our observations toward G+0.693 not only confirm the presence of  {\it s}-H$_2$C=CHOH in the ISM, but also present the first robust detection of the {\it anti} conformer of vinyl alcohol ({\it a}-H$_2$C=CHOH) in space (see Section$\,$\ref{others}).

Vinyl alcohol is a structural isomer of acetaldehyde (CH$_3$CHO) and therefore previous works have compared the abundance of both chemical species. \citet{turner01} showed that the column density of vinyl alcohol derived toward SgrB2(N) is similar to that measured for acetaldehyde. This is consistent with the results of \citet{agundez21} toward TMC-1, where the abundance ratio  {\it s}-H$_2$C=CHOH/CH$_3$CHO is $\sim$1. In contrast, \citet{martin-drumel19} did not detect vinyl alcohol in the EMoCA survey toward the high-mass star-forming region SgrB2(N2) \citep[][]{belloche16}, with an abundance ratio {\it s}-H$_2$C=CHOH/CH$_3$CHO$\leq$0.08.

Toward G+0.693, the measured abundance of CH$_3$CHO is 3.7$\times$10$^{-9}$ \citep[][]{sanz-novo22}, while the total abundance of vinyl alcohol (considering both the {\it syn} and {\it anti} forms) is $\sim$9$\times$10$^{-10}$. This implies that vinyl alcohol is a factor of $\sim$4 less abundant than CH$_3$CHO toward the G+0.693 Galactic Center molecular cloud, with a H$_2$C=CHOH/CH$_3$CHO abundance ratio of 0.25$\pm$0.02. This value lies between the abundance ratios found toward TMC-1 \citep[of 1.0$\pm$0.2;][]{agundez21} and SgrB2(N) \citep[of $\sim$1, although this value has large uncertainties; see][]{turner01}, and the one measured toward SgrB2(N2) \citep[of $\leq$0.08;][]{martin-drumel19}. The lack of detection of vinyl alcohol toward the SgrB2(N2) compact hot cores is consistent with the idea that the emission of vinyl alcohol arises mainly from the envelope of the SgrB2 molecular cloud. In Section$\,$\ref{vinyl:chem}, we discuss the possible chemical routes for the formation of vinyl alcohol.

\subsection{Chemistry of vinyl alcohol in G+0.693}
\label{vinyl:chem}

For vinyl alcohol, both gas-phase and grain-surface formation reactions have been proposed. For gas-phase formation, \citet{agundez21} concluded that the most likely mechanism for the formation of this species is the dissociative recombination of CH$_3$CHOH$^+$, for which the CCO backbone is formed with a braching ratio of 23\% \citep[][]{hamberg10}. The subsequent hydrogenation of CCO could form vinyl alcohol together with acetaldehyde \citep[][]{agundez21}, although the branching ratios are unknown. 
We also note that the gas-phase reactions OH + C$_2$H$_4$ and OH + CH$_2$CHCH$_3$ could also yield vinyl alcohol, although they present activation barriers \citep[][]{zhu05,zador09}. 

For grain-surface reactions, vinyl alcohol has been generated in interstellar ice analogues under proton irradiation of H$_2$O/C$_2$H$_2$ ices \citep[][]{hudson03} or electron irradiation of CO/CH$_4$ and H$_2$O/CH$_4$ ices \citep[][]{abplanalp15,bergantini17}. More recent experiments of non-energetic processing of ices also show that the co-deposition of C$_2$H$_2$ together with O$_2$ and H atoms induces the formation of vinyl alcohol and acetaldehyde under low temperature conditions \citep[T=10K; see][]{chuang20}. Unfortunately, these works do not provide information about the branching ratios for the two species, and therefore, it is not possible to assess which mechanism dominates the production of vinyl alcohol. 

For the case of the Galactic Center molecular cloud G+0.693, multiple routes (both gas-phase and grain-surface) may indeed be at play since its chemistry is dominated by shocks and high cosmic-ray ionization rates \citep[see e.g.][]{zeng18}. The kinetic temperatures of the molecular gas toward this source \citep[of 70-150$\,$K;][]{zeng18} could help overcome the activation barriers of the reactions OH + C$_2$H$_4$ and OH + CH$_2$CHCH$_3$ \citep[see above;][]{zhu05,zador09}. In addition, although the temperature of dust is low toward this cloud, the high radiation field of secondary UV photons \citep[induced by the high comic-rays ionization rate measured in the Galactic Center, which is factors 100-1000 higher than the standard value; see][]{goto14} could process the ices, generating enough radicals that could react {\it in-situ} on the surface forming vinyl alcohol. A detailed chemical modelling of the G+0.693 source is needed to asses all these different formation mechanisms. 

We finally note that the isomerization transformation of acetaldehyde into vinyl alcohol is not possible (either in the gas phase or in water ices) due to the large relative energies between both species and to the large activation energy barriers involved in this transformation \citep[see][]{perrero22}.

\subsection{Alcohols as precursors of lipids in primitive Earth}
\label{lipids}

Understanding the emergence of membranes is central to the problem of the origin of life. Membranes are indeed a key element since they enable the encapsulation and protection of the genetic information required for replication and of nutrients needed for metabolic processes \citep[][]{ruiz-mirazo14}. Current cell membranes are formed by a bilayer of phospholipids, i.e. amphiphilic molecules that contain a polar head and a nonpolar, hydrophobic tail. Phospholipids, however, present a rather complex structure with an alcohol phosphate group \citep[the polar hydrophilic head, where ethanolamine is one of the possible head groups;][]{rivilla21a}, a glycerol, and two hydrocarbon chains derived from fatty acids (the nonpolar hydrophobic tails). Because of this complexity, it is believed that these structures likely appeared biosynthetically at a later stage as a result of the cell's own machinery \citep[see][and references therein]{ruiz-mirazo14}. 

Alternatively, simpler amphiphilic molecules have been proposed as prebiotic lipids such as fatty alcohols or even prenols, i.e. single-chain unsaturated alcohols that present double bonds \citep[][]{ourisson95}. These amphiphilic compounds can assemble bi-layered supramolecular aggregates such as vesicles when they come in contact with water thanks to the so-called {\it hydrophobic effect}\footnote{The hydrophobic effect is the observed trend for nonpolar substances to aggregate and exclude water molecules in an aqueous solution.} and to van der Waals interactions. The discovery in the ISM of {\it n}-propanol (a small chain primary alcohol) and of vinyl alcohol (a small alcohol with a double bond; see Sections$\,$\ref{propanol} and \ref{others}) thus opens the possibility that these molecules were sourced into a primitive Earth through the impact of meteorites and comets, possibly representing precursors of prebiotic lipids such as fatty alcohols and prenols.  The fact that alcohols up to pentanol not only have been detected in comets such as 67P/Churyumov-Gerasimenko \citep[][]{altwegg19} but also in carbonaceous chondrites such as the Murray and Orgueil meteorites \citep[][]{jungclaus76,sephton02}, support the hypothesis of an exogenous delivery of precursors of prebiotic lipids into an early Earth.

\section{Conclusions}

In this paper we report the discovery of {\it n}-propanol in the ISM with the detection of the two lowest-energy conformers, {\it Ga}-{\it n}-C$_3$H$_7$OH and {\it Aa}-{\it n}-C$_3$H$_7$OH. {\it n}-propanol was detected toward the GMC G+0.693 located in the Galactic Center and known to be one of the richest repository of complex organic molecules in our Galaxy. The derived abundance of {\it n}-propanol (considering both $Ga$ and {\it Aa} conformers) is 6.6$\times$10$^{-10}$. By comparing this abundance with those of methanol and ethanol measured toward the same source, we find that the abundance of primary alcohols drops by approximately a factor of 10 by increasing complexity (i.e. by inserting a new CH$_2$ group). 

Finally, we also report the detection of both {\it syn} and {\it anti} conformers of vinyl alcohol (H$_2$C=CHOH) toward G+0.693. To our knowledge, this is the first time the latter conformer ({\it anti}) is firmly detected in the ISM. The derived abundance ratio between vinyl alcohol and acetaldehyde (its structural isomer) toward G+0.693 (of $\sim$0.25), lies between the ones found toward TMC-1 and the SgrB2(N) molecular envelope \citep[][]{turner01,agundez21}, and the upper limit measured toward the SgrB2(N2) compact hot core \citep[][]{martin-drumel19}. The detection of these alcohols toward G+0.693 opens the possibility for precursors of prebiotic lipids to form in the ISM. 

\begin{acknowledgements}
We would like to thank Carlos Briones for the fruitful discussions on possible prebiotic precursors of phospholipids and the formation of the first cell membranes. We also acknowledge the constructive and detailed report from an anonymous referee, who helped to improve the original version of the manuscript. Based on observations carried out with the IRAM 30m and Yebes 40m telescopes through projects 172-18 (PI: J. Mart\'{\i}n-Pintado), 018-19 (PI: V. M. Rivilla) and 20A008 (PI: I. Jim\'enez-Serra). IRAM is supported by INSU/CNRS (France), MPG (Germany) and IGN (Spain). The 40m radiotelescope at Yebes Observatory is operated by the Spanish Geographic Institute (IGN, Ministerio de Transportes, Movilidad y Agenda Urbana). I.J.-S. and J.M.-P. have received partial support from the Spanish State Research Agency (AEI) through project numbers PID2019-105552RB-C41 and MDM-2017-0737 Unidad de Excelencia $"$Mar\'{\i}a de Maeztu$"$ - Centro de Astrobiolog\'{\i}a (CSIC/INTA). L.F.R.-A. acknowledges support from a JAE-intro ICU studentship funded by the Spanish National Research Council (CSIC). L.F.R.-A., V.M.R. and L.C. also acknowledge support from the Comunidad de Madrid through the Atracci\'on
de Talento Investigador Modalidad 1 (Doctores con experiencia) Grant (COOL: Cosmic Origins Of Life; 2019-T1/TIC15379; PI: V.M.Rivilla). P.d.V. and B.T. thank the support from the European Research Council through Synergy Grant ERC-2013-SyG, G.A. 610256 (NANOCOSMOS) and from the Spanish Ministerio de Ciencia e Innovación (MICIU) through project PID2019-107115GB-C21. B.T. also acknowledges the Spanish MICIU for funding support from grant PID2019-106235GB-I00.

\end{acknowledgements}

%
%

\begin{appendix} 

\section{Rotational Diagrams}
\label{rotdiag}

\begin{figure}
   \centering
   \includegraphics[angle=0,width=0.47\textwidth]{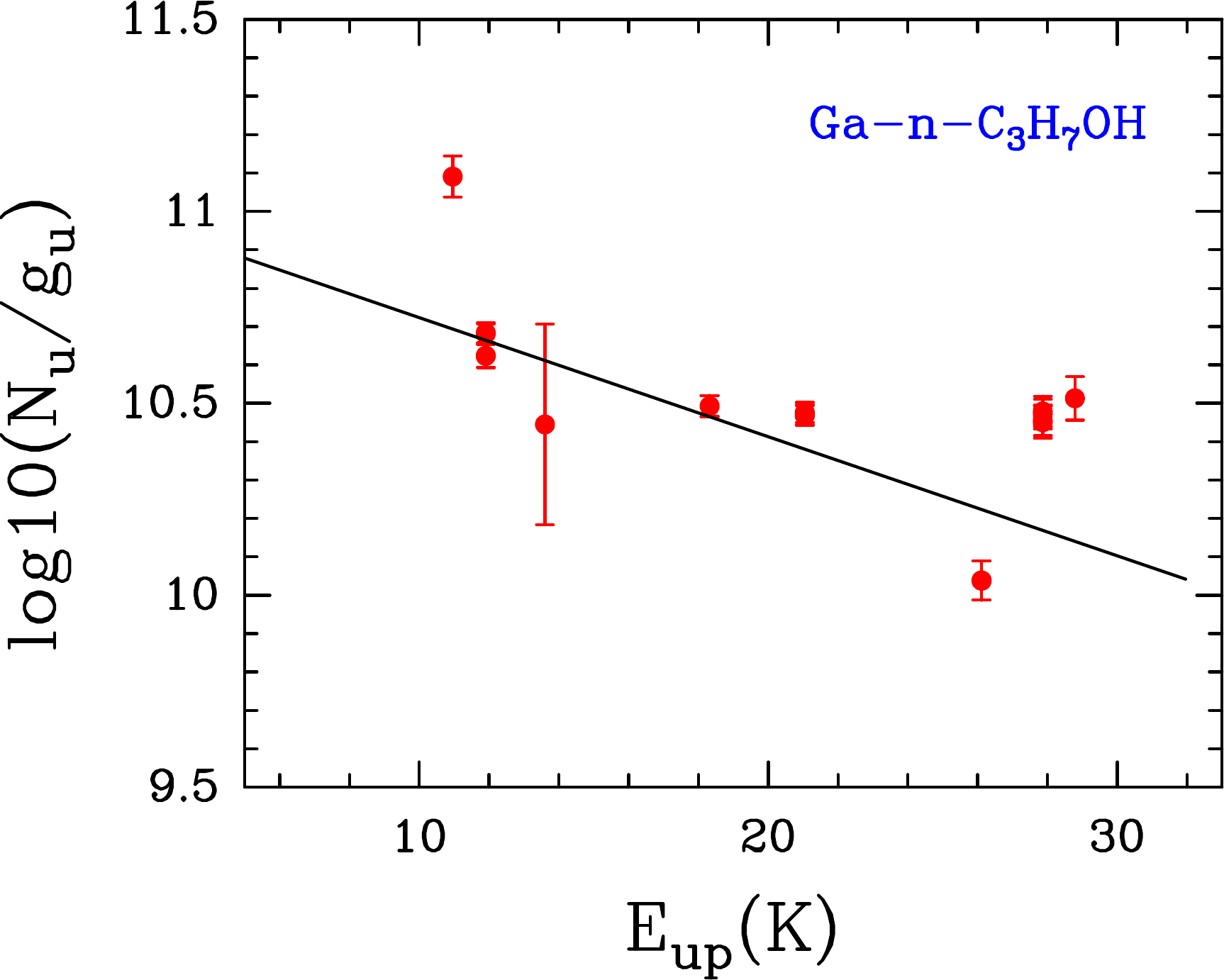}
      \caption{Rotational diagram obtained by using the rotational transitions of $Ga$-$n$-C$_3$H$_7$OH reported in Table$\,$\ref{tab:propanol}. The derived parameters are T$_{ex}$=16.9$\pm$3.1$\,$K  and N=6.6$\pm$2.3$\times$10$^{13}$$\,$cm$^{-2}$.\label{fig:rotdiag-Ga}}
   \end{figure}

\begin{figure}
   \centering
   \includegraphics[angle=0,width=0.47\textwidth]{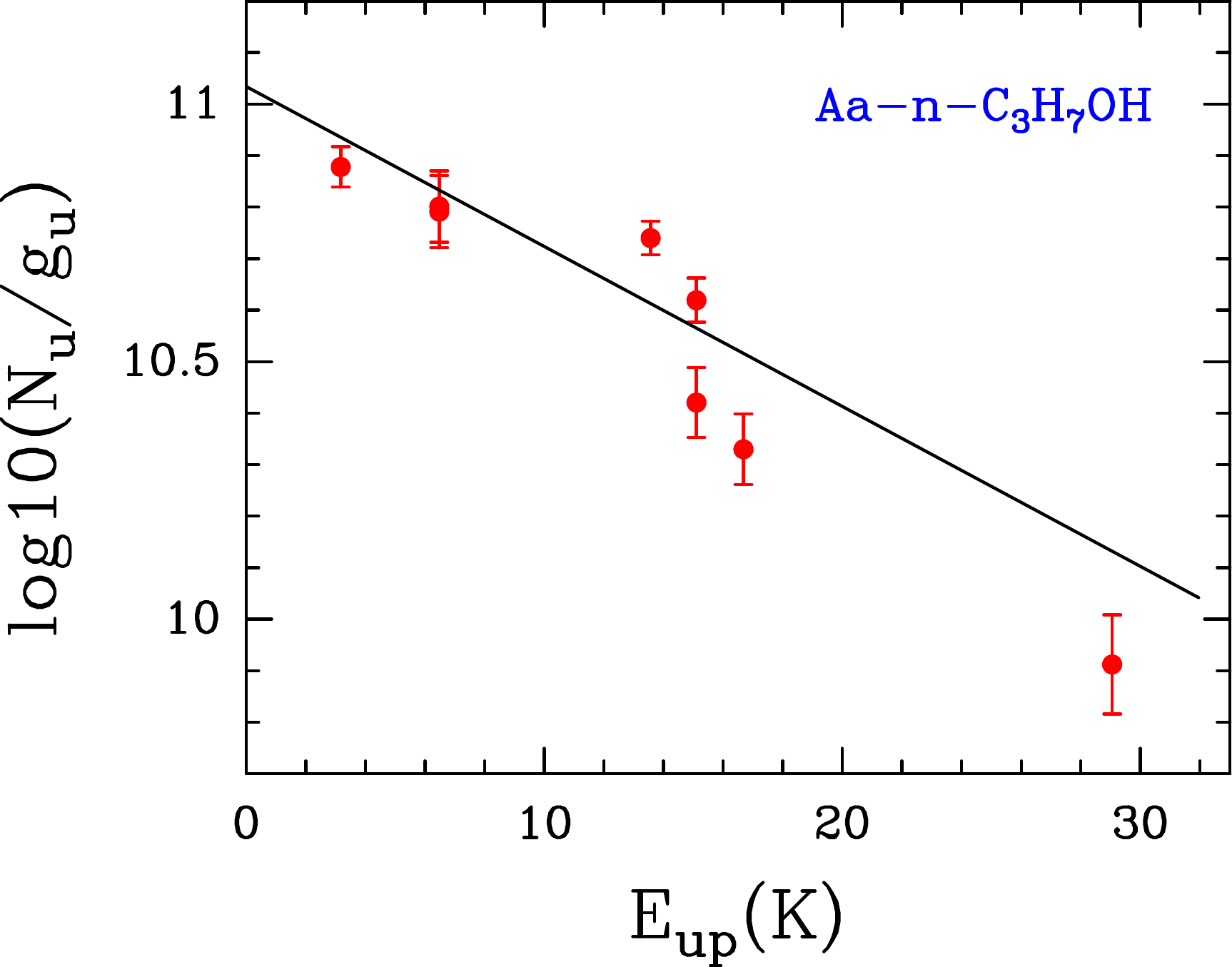}
      \caption{Rotational diagram obtained by using the rotational transitions of $Aa$-$n$-C$_3$H$_7$OH reported in Table$\,$\ref{tab:propanol}, except the 9$_{2,8}$$\rightarrow$9$_{1,9}$ and 10$_{1,10}$$\rightarrow$9$_{0,9}$ lines that are blended with U species. The derived parameters are T$_{ex}$=14.0$\pm$3.4$\,$K  and N=5.1$\pm$2.2$\times$10$^{13}$$\,$cm$^{-2}$. \label{fig:rotdiag-Aa}}
   \end{figure}

\begin{figure}
   \centering
   \includegraphics[angle=0,width=0.47\textwidth]{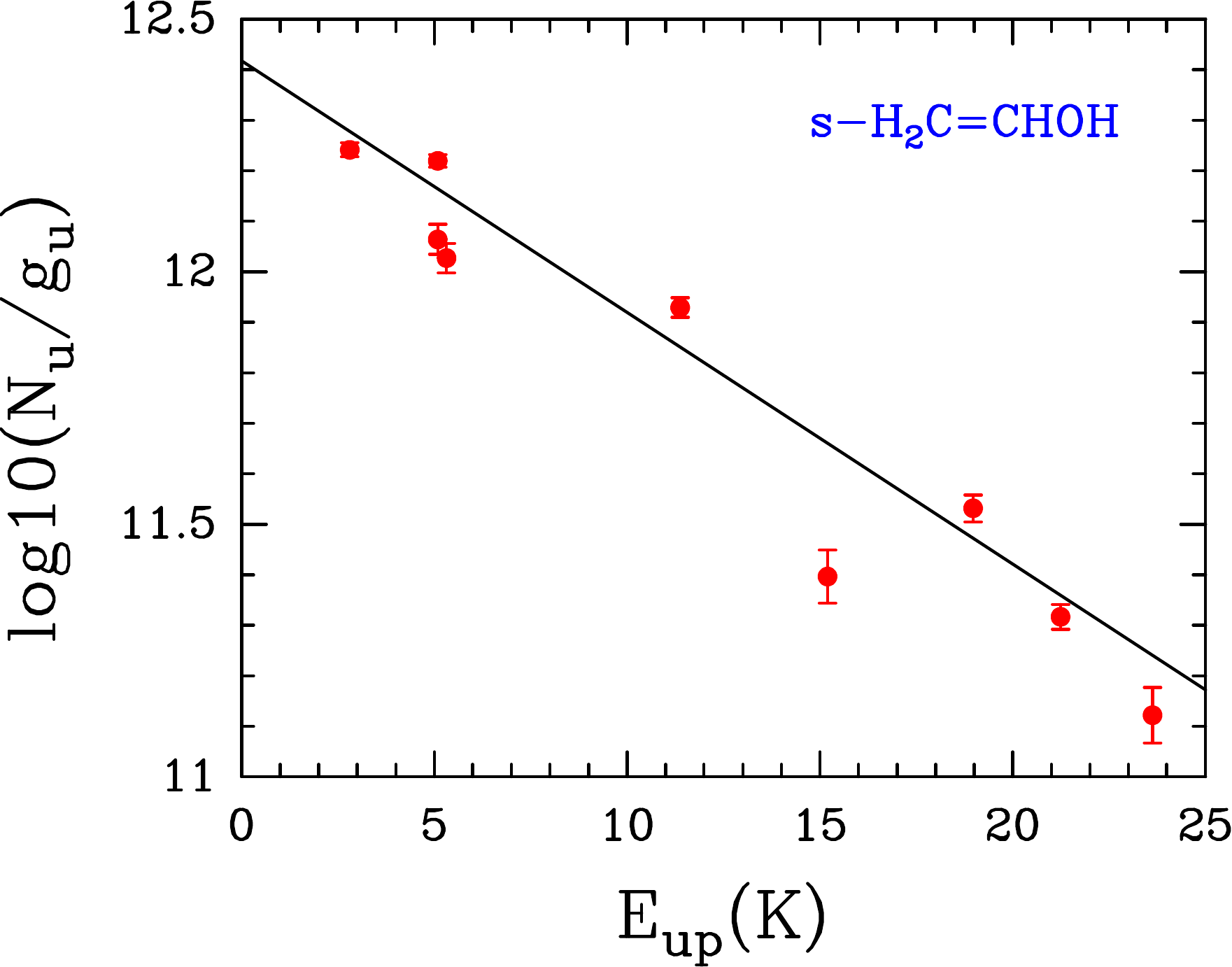}
      \caption{Rotational diagram obtained by using the rotational transitions of $s$-H$_2$C=CHOH reported in Table$\,$\ref{tab:vinylalcohol}. The derived parameters are T$_{ex}$=8.715$\pm$0.008$\,$K  and N=1.5$\pm$0.3$\times$10$^{14}$$\,$cm$^{-2}$.\label{fig:rotdiag-s}}
   \end{figure}

\begin{figure}
   \centering
   \includegraphics[angle=0,width=0.47\textwidth]{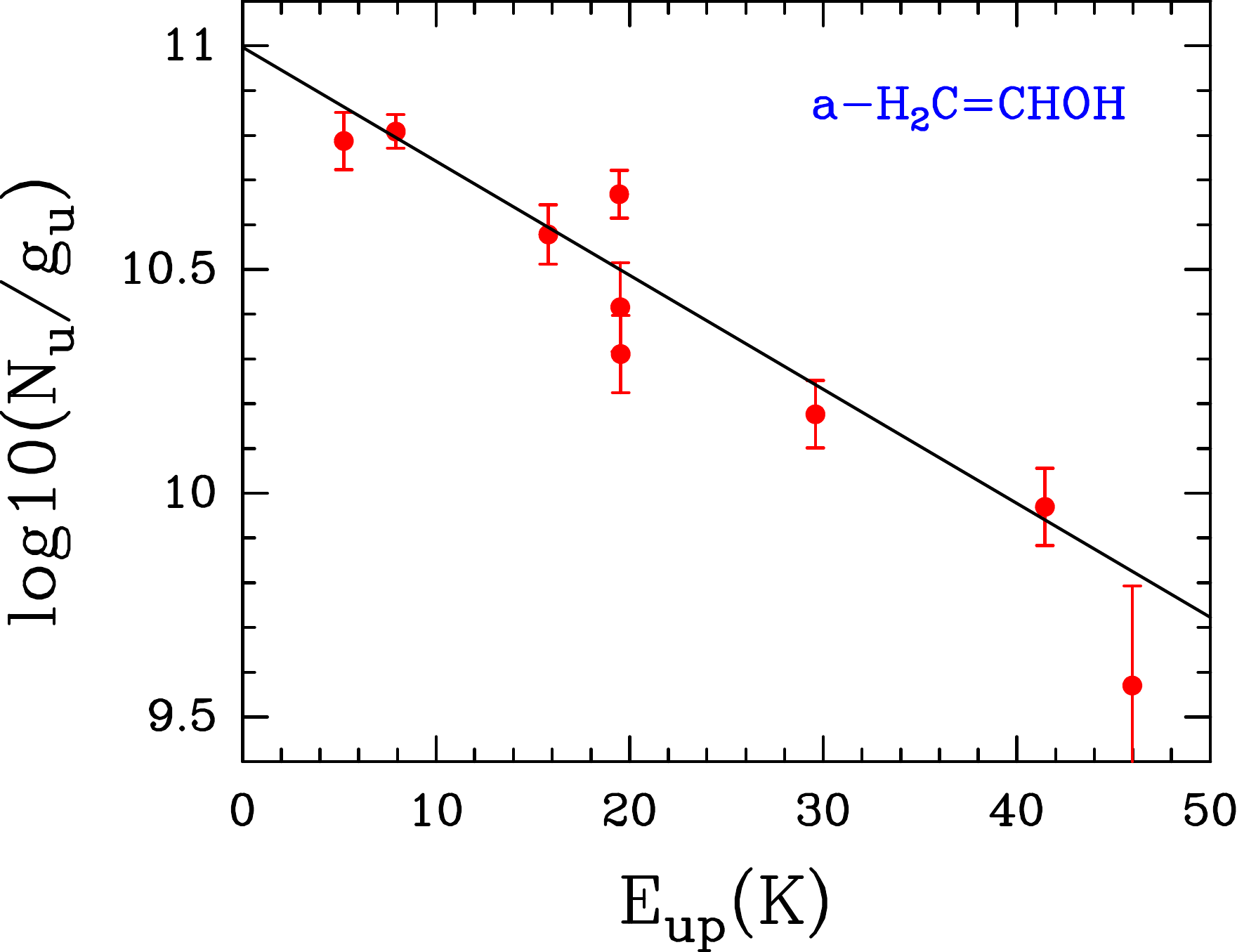}
      \caption{Rotational diagram obtained by using the rotational transitions of $a$-H$_2$C=CHOH reported in Table$\,$\ref{tab:vinylalcohol}. The derived parameters are T$_{ex}$=17.0$\pm$2.4$\,$K  and N=1.5$\pm$0.4$\times$10$^{13}$$\,$cm$^{-2}$.\label{fig:rotdiag-a}}
   \end{figure}

In this Section, we show the rotational diagrams obtained for $Ga$-$n$-C$_3$H$_7$OH, $Aa$-$n$-C$_3$H$_7$OH, $s$-H$_2$C=CHOH and $a$-H$_2$C=CHOH (see Figures$\,$\ref{fig:rotdiag-Ga}, \ref{fig:rotdiag-Aa}, \ref{fig:rotdiag-s}, and \ref{fig:rotdiag-a}), which have been detected toward G+0.693. The goal is to establish the reliability of the excitation temperatures (T$_{ex}$) and column densities (N) derived by Madcuba (see Sections$\,$\ref{propanol} and \ref{others}). For these diagrams, we have used all molecular transitions reported in Tables$\,$\ref{tab:propanol} and \ref{tab:vinylalcohol}, except for  $Aa$-$n$-C$_3$H$_7$OH for which we have not considered the 9$_{2,8}$$\rightarrow$9$_{1,9}$ and 10$_{1,10}$$\rightarrow$9$_{0,9}$ transitions that appear blended with U species. 

The results of the rotational diagrams are: T$_{ex}$=16.9$\pm$3.1$\,$K  and N=6.6$\pm$2.3$\times$10$^{13}$$\,$cm$^{-2}$ for $Ga$-$n$-C$_3$H$_7$OH; T$_{ex}$=14.0$\pm$3.4$\,$K  and N=5.1$\pm$2.2$\times$10$^{13}$$\,$cm$^{-2}$ for $Aa$-$n$-C$_3$H$_7$OH; T$_{ex}$=8.715$\pm$0.008$\,$K  and N=1.5$\pm$0.3$\times$10$^{14}$$\,$cm$^{-2}$ for $s$-H$_2$C=CHOH; and T$_{ex}$=17.0$\pm$2.4$\,$K  and N=1.5$\pm$0.4$\times$10$^{13}$$\,$cm$^{-2}$ for $a$-H$_2$C=CHOH. These values are consistent with those obtained by Madcuba. 

\end{appendix}

\end{document}